\documentclass[twoside,twocolumn,9pt]{article}
\usepackage{amssymb} 
\usepackage{extsizes}
\usepackage[super,sort&compress,comma]{natbib} 
\usepackage[version=3]{mhchem}
\usepackage[left=1.5cm, right=1.5cm, top=1.785cm, bottom=2.0cm]{geometry}
\usepackage{balance}
\usepackage{mathptmx}
\usepackage{sectsty}
\usepackage{graphicx} 
\usepackage{lastpage}
\usepackage[format=plain,justification=justified,singlelinecheck=false,font={stretch=1.125,small,sf},labelfont=bf,labelsep=space]{caption}
\usepackage{float}
\usepackage{fancyhdr}
\usepackage{fnpos}
\usepackage[english]{babel}
\addto{\captionsenglish}{%
 
}
\usepackage{array}
\usepackage{droidsans}
\usepackage{charter}
\usepackage[T1]{fontenc}
\usepackage[usenames,dvipsnames]{xcolor}
\usepackage{setspace}
\usepackage[compact]{titlesec}
\usepackage{hyperref}

\usepackage{epstopdf}

\definecolor{cream}{RGB}{222,217,201}
\begin{document}

\pagestyle{fancy}
\thispagestyle{plain}
\fancypagestyle{plain}{
\renewcommand{\headrulewidth}{0pt}
}

\makeFNbottom
\makeatletter
\renewcommand\LARGE{\@setfontsize\LARGE{15pt}{17}}
\renewcommand\Large{\@setfontsize\Large{12pt}{14}}
\renewcommand\large{\@setfontsize\large{10pt}{12}}
\renewcommand\footnotesize{\@setfontsize\footnotesize{7pt}{10}}
\makeatother

\renewcommand{\thefootnote}{\fnsymbol{footnote}}
\renewcommand\footnoterule{\vspace*{1pt}%
\color{cream}\hrule width 3.5in height 0.4pt \color{black}\vspace*{5pt}} 
\setcounter{secnumdepth}{5}

\makeatletter 
\renewcommand\@biblabel[1]{#1}      
\renewcommand\@makefntext[1]%
{\noindent\makebox[0pt][r]{\@thefnmark\,}#1}
\makeatother 
\renewcommand{\figurename}{\small{Fig.}~}
\sectionfont{\sffamily\Large}
\subsectionfont{\normalsize}
\subsubsectionfont{\bf}
\setstretch{1.125} 
\setlength{\skip\footins}{0.8cm}
\setlength{\footnotesep}{0.25cm}
\setlength{\jot}{10pt}
\titlespacing*{\section}{0pt}{4pt}{4pt}
\titlespacing*{\subsection}{0pt}{15pt}{1pt}

\fancyfoot{}
\fancyfoot[LO,RE]{\vspace{-7.1pt}\includegraphics[height=9pt]{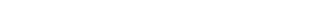}}
\fancyfoot[CO]{\vspace{-7.1pt}\hspace{13.2cm}\includegraphics{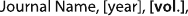}}
\fancyfoot[CE]{\vspace{-7.2pt}\hspace{-14.2cm}\includegraphics{head_foot/RF}}
\fancyfoot[RO]{\footnotesize{\sffamily{1--\pageref{LastPage} ~\textbar \hspace{2pt}\thepage}}}
\fancyfoot[LE]{\footnotesize{\sffamily{\thepage~\textbar\hspace{3.45cm} 1--\pageref{LastPage}}}}
\fancyhead{}
\renewcommand{\headrulewidth}{0pt} 
\renewcommand{\footrulewidth}{0pt}
\setlength{\arrayrulewidth}{1pt}
\setlength{\columnsep}{6.5mm}
\setlength\bibsep{1pt}

\makeatletter 
\newlength{\figrulesep} 
\setlength{\figrulesep}{0.5\textfloatsep} 

\newcommand{\topfigrule}{\vspace*{-1pt}%
\noindent{\color{cream}\rule[-\figrulesep]{\columnwidth}{1.5pt}} }

\newcommand{\botfigrule}{\vspace*{-2pt}%
\noindent{\color{cream}\rule[\figrulesep]{\columnwidth}{1.5pt}} }

\newcommand{\dblfigrule}{\vspace*{-1pt}%
\noindent{\color{cream}\rule[-\figrulesep]{\textwidth}{1.5pt}} }

\makeatother

\twocolumn[
 \begin{@twocolumnfalse}
{\includegraphics[height=30pt]{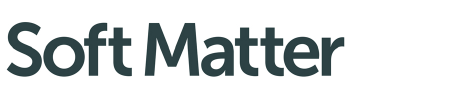}\hfill\raisebox{0pt}[0pt][0pt]{\includegraphics[height=55pt]{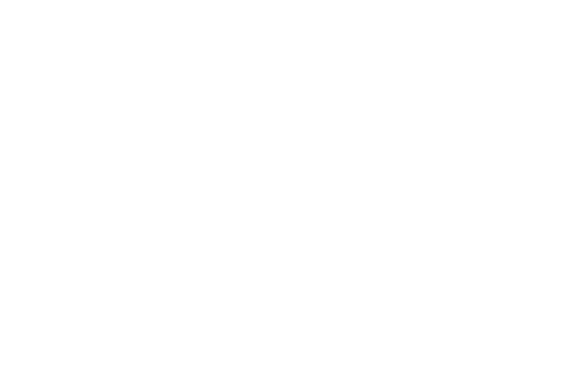}}\\[1ex]
\includegraphics[width=18.5cm]{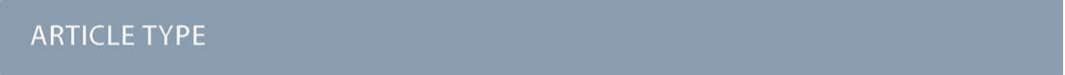}}\par
\vspace{1em}
\sffamily
\begin{tabular}{m{4.5cm} p{13.5cm} }

\includegraphics{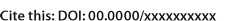} & \noindent\LARGE{\textbf{Faceted wrinkling by contracting a curved boundary$^\dag$}} \\
\vspace{0.3cm} & \vspace{0.3cm} \\

 & \noindent\large{Anshuman S. Pal,\textit{$^{a}$} Luka Pocivavsek,\textit{$^{b}$} and Thomas A. Witten\textit{$^{a}$}$^{\ast}$} \\

\includegraphics{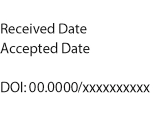} & \noindent\normalsize{
Single-mode deformations of two-dimensional materials, such as the Miura-\textit{ori} zig-zag fold, are important to the design of deployable structures because of their robustness; these usually require careful pre-patterning of the material. 
Here we show that inward contraction of a curved boundary produces a fine wrinkle pattern with a novel structure that suggests similar single-mode characteristics, but with minimal pre-patterning.
Using finite-element representation of the contraction of a thin circular annular sheet, we show that these sheets wrinkle into a structure well approximated by an isometric structure composed of conical sectors and flat, triangular facets.
Isometry favours the restriction of such deformations to a robust low-bending energy channel that avoids stretching. This class of buckling offers a novel way to manipulate sheet morphology via boundary forces. 

} \\

\end{tabular}

 \end{@twocolumnfalse} \vspace{0.6cm}

 ]

\renewcommand*\rmdefault{bch}\normalfont\upshape
\rmfamily
\section*{}
\vspace{-1cm}


\footnotetext{\textit{$^{a}$~James Franck Institute and Dept.\ of Physics, University of Chicago, IL, USA.}}
\footnotetext{\textit{$^{b}$~Dept. of Surgery, Pritzker School of Medicine, University of Chicago, Chicago, IL, USA.}}
\footnotetext{$^{\ast}$~E-mail: tten@uchicago.edu}

\footnotetext{\dag~Electronic Supplementary Information (ESI) available: [details of any supplementary information available should be included here]. See DOI: 10.1039/cXsm00000x/}



\begin{figure*}[htb]
\centering
\includegraphics[width=0.7\textwidth]{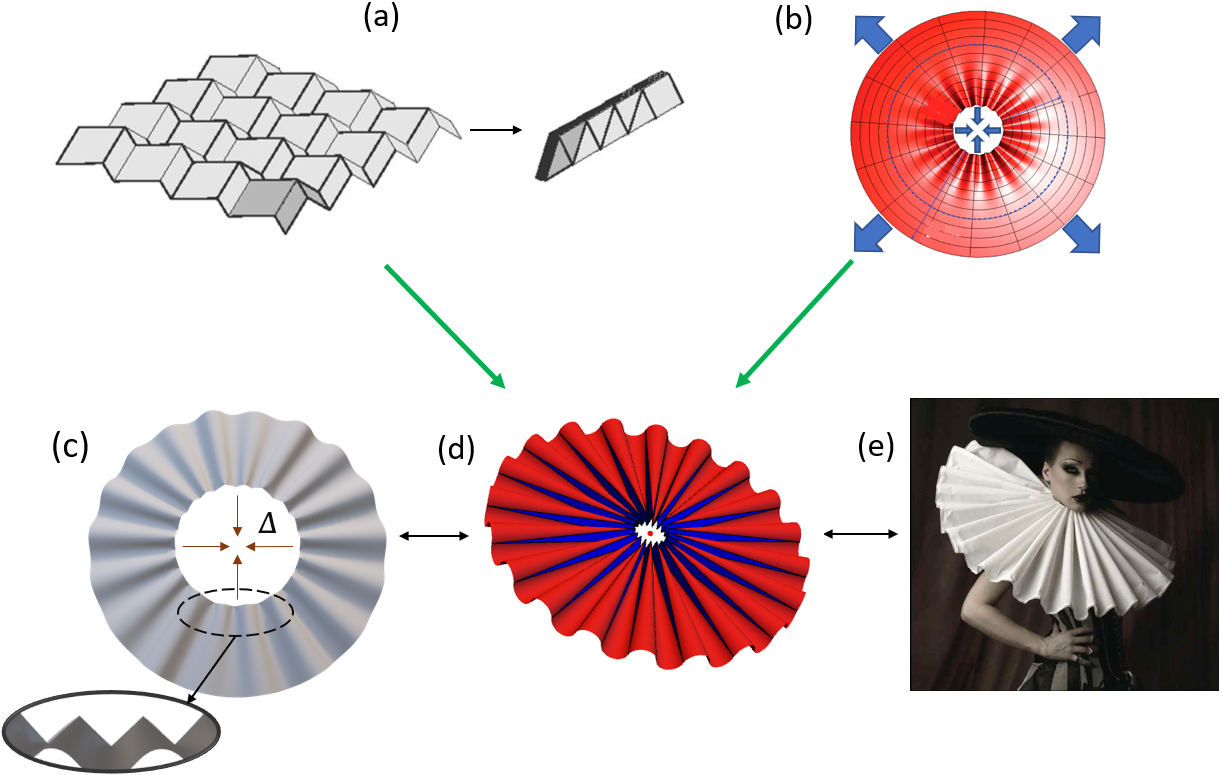}
\caption{Geometry-controlled robust radial wrinkling. a) Miura-\textit{ori} patterns guide a flat sheet to contract into a folded state with a single, continuous motion (adapted from \cite{Schenk2013}). b) Smooth radial (i.e.,~Lam\'e) wrinkling in an annulus under both inner and outer tension, indicated by arrows (adapted from \cite{Davidovitch2011}). 
In this paper, we study Miura-\textit{ori} like robust deformation of an annulus in a radial Lam\'e setting. This deformation can be generated in multiple ways. c) Finite-element calculated shape under uniform radial contraction of the inner boundary, described in \textit{Methods} (inset shows detail of emergent zigzag inner boundary). d) Circular-cone-triangle geometric construction defined in Sec.~\ref{sec:constructionSolution}, with triangles shown in blue and cones in red. e) Ruff collar garment (reproduced from \cite{ruffCollarPinterest}) made by manually pleating the inner boundary of an annular piece of cloth.}
\label{fig:1}
\end{figure*}
The flourishing field of extreme mechanics \cite{Krieger2012, Bertoldi2017, Santangelo2017,Callens2018} seeks ways in which large, spontaneous material deformations yield robust, reproducible structures and motions. Many of these structures exploit the selective deformations shown by thin sheets, which bend or fold easily but resist internal stretching strongly \cite{Cerda1998, Cerda2003, Cerda2004, mora_buckling_2006, muller_conical_2008, Klein2007, Vandeparre2011, Hure2012, Vella2015, Davidovitch2019, Paulsen2019, Efrati2015}. The large resistance to stretching creates strong, non-local constraints that restrict the sheet to selected modes of deformation. A classic example is the Miura-\textit{ori} folding pattern \cite{Miura1985, Mahadevan2005, Schenk2013}. This periodic pattern enables one to contract a spread-out flat sheet of paper into a compactly folded state with one single mode of deformation dictated by the pattern (see Fig.~\ref{fig:1}a). Here we report similar robust deformation of a flat sheet without the position-dependent processing needed for an origami fold. Instead, we specify an initial annular shape and a simple mode of \textit{edge} deformation: contracting the inner boundary radially. The result is a strongly reproducible three-dimensional pattern of high-amplitude wrinkles. We attribute the reproducibility to a qualitative feature of the simulated, finite-element simulation---each wrinkle approaches an isometric shape, with no stretching. This unstretched morphology is made possible by a distinct pattern of flat, triangular facets alternating with smoothly-curving conical segments.

We consider a minimal form of radial contraction that we call ``inner Lam\'e" contraction. We displace the inner boundary of the annulus with no further constraints. Our interest in this geometry arose from experiments on the curling of edge-connected polymeric sheets with unequal solvent swelling. Ref.~\cite{Kim2012Thermally-respo} shows the result for two half sheets with different swelling separated by a straight boundary. We simulated the radial analogue of this situation, where the straight boundary was replaced by a circular boundary separating an inner region with little swelling from an outer one with more swelling. The swelling resulted in fine radial wrinkling like that shown in Fig.~\ref{fig:1}. The present simplified system was developed to understand this wrinkling.  
 
As shown in Fig.~\ref{fig:1}c, the inner Lam\'e  contraction produces fine wrinkling like the conventional Lam\'e contraction of Fig.~\ref{fig:1}b. However the two wrinkled states are radically different. The conventional Lam\'e wrinkle wavelength is explained in terms of the applied tension \cite{Davidovitch2011}. However, in the inner Lam\'e, the exterior tension of the Lam\'e state is absent, and so the established mechanism for the wrinkle wavelength cannot apply. Indeed, our wrinkling contrasts with most wrinkled systems, where the bending energy favoring longer wavelengths is countered by the energy associated with some external forcing \cite{Cerda2003,brau_wrinkle_2013}. The wrinkle wavelength then depends explicitly on the strength of this forcing. The inner Lam\'e system has no such strength parameter. Instead, the deformation is defined in terms of the geometric displacement of the inner edge only. 

Even among geometrically-defined forms of wrinkling, the inner-Lam\'e geometry is anomalous.\textit{ A priori}, it doesn't conform to typical one-dimensional buckling treatable as an Euler elastica system \cite{Levien:2008vf}. Other forms in which the constraint is applied at one boundary \cite{Cerda2005, Vandeparre2011} do not give a well-defined wave number as we find.

Furthermore, we note that the forms of wrinkling mentioned above are explained as equilibrium ground-states of the system. In our system the wrinkles emerge from a quasistatic deformation. That is, the inner displacement is applied gradually, and the configuration is allowed to relax at every stage. This procedure corresponds to how one would perform the deformation experimentally. In this limit, the system follows a given energy minimum as the deformation is imposed. It need not be an energy ground state.

In this paper, we address the structure of the individual wrinkles resulting from this unusual mode of deformation\footnote{For now, we defer the obvious issue of accounting for the observed wrinkle wavelength. Understanding the wrinkle structure seems prerequisite to accounting for the wavelength.
}.
We show that a strain-free isometric shape quantitatively accounts for observed features of our simulated shape, notably the elastic energy shown in Fig.~\ref{fig:5}. This energy contrasts strongly with the much higher energy expected with a simple sinusoidal wrinkling pattern (Sec. \ref{sec:results_energyComparison}).
Moreover, the simulated deformation matches an explicitly isometric deformation of the original annulus into a cone-triangle pattern. We describe the robustness of the structure using a variety of initial and boundary conditions for the contracted inner boundary. We argue that the observed robustness of the cone-triangle structure is a natural consequence of its isometric property. We note how the cone-triangle structure can be induced via different paths, as illustrated in Figs.~\ref{fig:1}c, d, and e, and discuss generalisations to a broader range of structures. Using the findings of this paper, we address the question of wavelength determination in a separate work \cite{Pal2023}. 

Sections \ref{sec:innerLame} and \ref{sec:constructionSolution} define the inner Lam\'e deformation that we simulate, and the isometric model shape used for comparison. In Section \ref{sec:results}, we show that the shape and energy of the simulated sheet are consistent with the model. Section \ref{sec:discussion} addresses our assumptions and discusses the value of such deformations for generating spontaneous structures. Finally, Section \ref{sec:methods} gives an account of our numerical methods.

\section{Inner Lam\'e deformation} \label{sec:innerLame}
\begin{figure}[htb]
\centering
\includegraphics[width=0.48\textwidth]{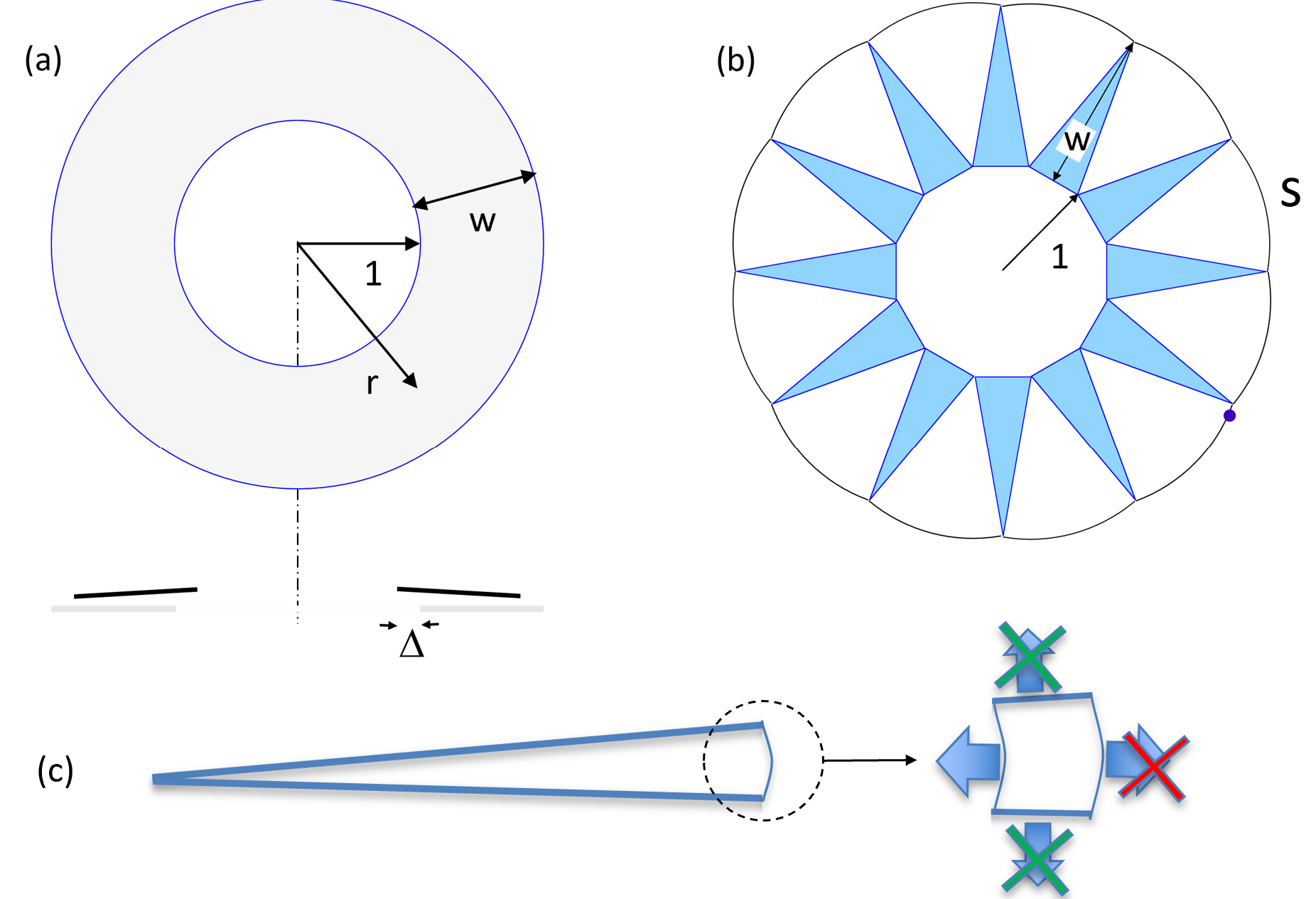}
\caption{a) Sketch of the inner Lam\'e deformation of a horizontal annulus. (Top) Top view showing the undeformed annulus, of inner radius $1$ and width $w$. (Bottom) Vertical cross-section of the deformed annulus through a line where the vertical displacement of the wrinkle is maximal. The undeformed state is shown in light shading; the dark lines are a section of the deformed annulus, contracted horizontally by a distance $\Delta$. These lines are also displaced upward and tilted downward. b) Sketch of the circular-cone-triangle construction in its initial, flat state. Here we choose wavenumber $m=6$, hence it has 12 triangles (shown in colour). The mid-line and height $w$ of one triangle are indicated. The sectors bridging the triangles are shown in white; $S$ is its outermost arc length.  When the bases of the triangles are moved inward, they must tilt to retain their shape and connectivity. The bridging sectors are squeezed laterally to form cones. A typical deformed example is shown in Fig.~\ref{fig:1}c.
c) Top view of a narrow sector of the sheet, with detail of an area element at the outer edge of Lam\'e sheet showing radial (horizontal) and azimuthal (almost vertical) boundary forces. In the fully buckled (FT) regime \cite{Davidovitch2011}, the azimuthal forces nearly vanish (as indicated by the upper and lower X's). In the the present ``inner Lam\'e" system, the outer radial force also vanishes, indicated by the X at right.
Thus, the inward radial force must also nearly vanish for force equilibrium. All elements to the left of the pictured one are subject to the same argument. Thus except for bending stresses, all radial stresses vanish.
}
\label{fig:2}
\end{figure}

Given the circular annulus of Fig.~\ref{fig:2}a, we define the inner Lam\'e deformation as the equilibrium shape induced by drawing the inner boundary inward by a distance $\Delta$, so that it is forced to live on a cylinder whose radius is reduced by $\Delta$.\footnote{As discussed in Sec.~\ref{sec:discussion}, we constrain the inner boundary to move only parallel to the axis of the cylinder.} We define the initial inner radius as our unit of length. The pull distance $\Delta$ thus acts as the sole loading parameter in the system. The system can thus be conveniently defined using only three geometric parameters: thickness $t$, the width $w$, and displacement `loading' $\Delta$. The internal forces determining the shape arise from the bending modulus $B$ and the in-plane stretching modulus $Y$. Here, we seek the asymptotic behaviour of our annulus in the thin limit: $t \ll w; t \ll 1$.

The conventional ``FT" reasoning used to explain classical Lam\'e wrinkling \cite{Davidovitch2011} suggests that fine wrinkling like that of Fig. \ref{fig:1}.  would require negligible radial stretching. In the FT regime, the wrinkling slopes grow to the order of unity, and the associated azimuthal stress becomes negligible compared to the pre-buckling stress.
Only the \textit{bending azimuthal stress}, owing to wrinkling, proportional to the bending modulus and the curvature, remains. The same is expected for our inner Lam\'e case: consider the radial stress in a narrow sector of the annulus (see Fig.~\ref{fig:2}c). Equilibrium requires that the net radial force on the sector be zero. In the unbuckled state, the inner tension driving the deformation is balanced by the net outward component of the azimuthal stresses on the sides of the sector. But in the FT regime, this stress is negligible, as argued above. In addition, there is no radial tension at the outer boundary. We conclude that the only forces to balance an inner tension are the weak bending stresses.

The absence of any deformation stress except for bending stress would imply that the sheet deforms only by bending in the thin limit. That is, it is isometric to the undeformed flat annulus. Such an isometric limit can be attained by wrinkles of diverging fineness \cite{Vella2015, Davidovitch2019}. But if a limiting smooth shape is to exist, then it must obey the geometrical property of ``developability". That is, in view of Gauss's Theorema Egregium \cite{doCarmo}, there must be a straight generator line through each point of the surface that extends to the boundary without curving in space \cite{Venkataramani:2000cr}. In the next sections, we show that the contracted annulus is indeed consistent with a specific developable surface, suggested by the numerically determined shape.
\section{A developable cone-triangle model} \label{sec:constructionSolution}

 The faceted inner Lam\'e morphology of Fig.~\ref{fig:1} suggests a further hypothesis: that the wrinkled morphology can be seen as a piecewise union of alternating flat triangles and curved conical sectors. We take this as the starting point for our geometric model. For a configuration with wavenumber $m$, we consider a flat configuration (see Fig.~\ref{fig:2}b) where the inner boundary is a $2m$-sided polygon. Each segment of this boundary is the base of an isosceles triangle of height $w$. To complete the annulus, the apexes of adjacent triangles are joined with an arc. The constructed surface then consists of these triangles and the sectors joining them.

One may readily specify an isometric deformation that allows the inner boundary of this structure to contract. We deform the initial surface by translating the base of each triangle inward toward the centre, specifying that the triangles remain rigid. In order to maintain their shape and connectivity, the triangles must tilt (i.e.~rotate about their mid-lines) increasingly as they are drawn inward; adjacent triangles tilt in opposite directions. The sectors connecting two triangles must then also bend in order to remain connected to their triangles. A simple \textit{isometric} choice for this deformation is to bend the flat sectors into \textit{circular conical arcs} as the boundaries of each sector squeeze together. These arcs do not in general join smoothly with their adjacent triangles, as further discussed in Sec.~\ref{sec:results_energyComparison}.
\section{Comparing simulated deformation with model} \label{sec:results}
In this section, we test whether the isometric/developable cone-triangle construction is able to predict the quantitative features of our simulated inner Lam\'e deformation from Sec.~\ref{sec:innerLame}. Our finite-element analysis (FEA) software Abaqus uses standard methods to represent the deformed equilibrium shape as the (cylindrical) inner boundary is gradually contracted inward. The inner boundary is allowed to buckle by moving along the cylinder (see \textit{Methods}). For simplicity we restrict this motion to the axial direction, and allow no azimuthal motion of the boundary points. We describe the effect of this restriction in Sec.~\ref{sec:lateralConstraint}. 

In comparing this calculation to the cone-triangle construction, we must supply two free parameters: the boundary displacement $\Delta$ and the wavenumber $m$. Thus, we record the $\Delta$ and $m$ of the finite-element deformation, and then construct the corresponding geometric solution with the same $\Delta$ and $m$. 
\begin{figure*}[htb]\centering
\includegraphics[width=0.9\textwidth]{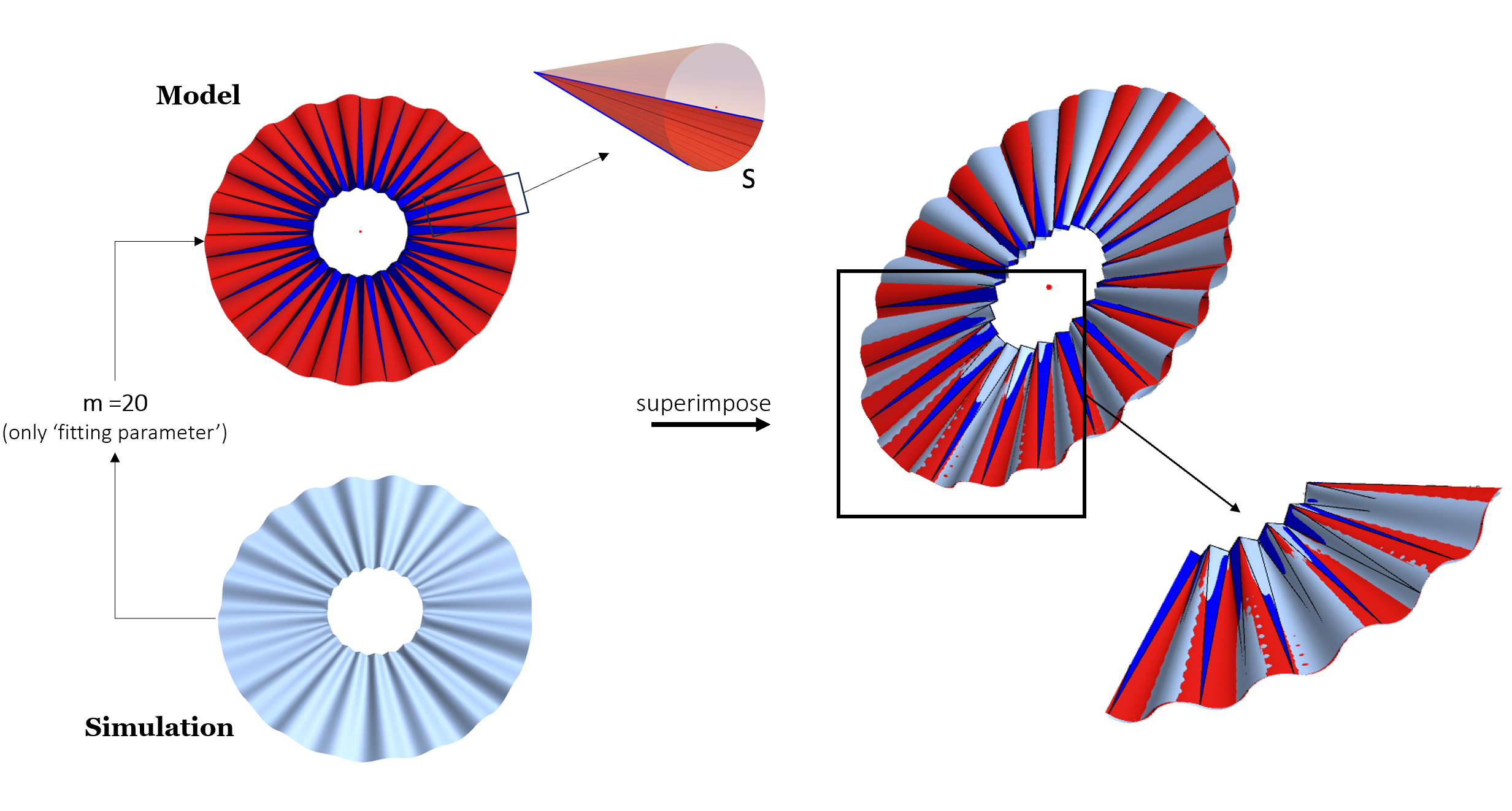}
\caption{Qualitative comparison between simulations and our cone-triangle model for a representative sample ($w=1.67$, $t=1.33\times 10^{-3}$). (Left) Given $w$, $\Delta$ and the wavenumber $m$ from the numerical solution, we construct a model cone-triangle solution (cones in red, triangles in blue) using the procedure described in Section \ref{sec:constructionSolution}. (inset) Zoom-in of a single conical sector (in red) with the entire cone shown (in cream). $S$ marks the length of the outermost arc (as in Fig.~\ref{fig:2}b) after deformation. For details of construction, see SI. (Right) For a qualitative comparison, we superpose the two solutions, with azimuthal rotation as the only degree of freedom. We find close agreement between them; with this top-down view, areas with colour indicate where the model solution is above the numerical solution, while those in grey indicate the opposite. In particular, the zoomed-in boxed portion shows close matching at both inner and outer boundaries.  Specifically, the two inner boundaries show a regular zigzag shape of the same amplitude, which extends to qualitatively flat facets and terminates in a smoothly undulating outer boundary that also has similar amplitude in the two sheets.
 Zooming out to the whole annulus, we see that there is a ‘phase mismatch’ between the two solutions, with part of the solution in-phase (the boxed region) and part of it out-of-phase. This is due to the non-uniform size of the wrinkles in the numerical solution.
}
\label{fig:3}
\end{figure*}
The first comparison we make is purely qualitative -- we superpose the two solutions and observe how well they match up. Fig.~\ref{fig:3} shows such a comparison, where we have rotated the geometric solution about the cylinder axis to obtain the best superposition. As the figure shows, one may align the two figures so that they overlay well over a significant part of the annulus, with matching zigzags at the inner boundary and matching waves at the outer boundary. Between the inner and outer boundaries, the surfaces match closely enough that the two surfaces alternately hide each other. 

The qualitative resemblance seen in the figure encourages a more quantitative comparison. In particular, we propose that for a given wave number $m$, the shape of the sheet approaches an isometric cone-triangle structure in the limit of zero thickness. Further, approximating each cone as a circular cone sector that connects its neighboring triangles gives a useful match to the shape and energy of the sheet. 
\par
Accurate comparison is hindered, however, by the variability of the simulated wrinkles. Though regular in general, the simulated wrinkle pattern is not strictly periodic on the annulus. To optimise the comparison, we perturb the initial flat state of the numerical sheet to promote periodicity. Specifically, we deform the initial state with a low-amplitude, smooth, periodic wrinkle with a wavelength similar to the original simulations. \footnote{The initial wrinkle shape chosen is an eigenmode obtained from linear stability analysis performed using the Abaqus software.}

\subsection{Inextensibility and bending strains}\label{sec:results_inextensibility}
We may roughly verify the anticipated isometry of the sheet as announced in the \textit{Introduction}. 
To this end, we measure changes in lengths of the material circles at the inner ($r=1$) and outer ($r=1+w$) boundaries.

\begin{figure*}[htb]
\centering
\includegraphics[width=0.95\textwidth]{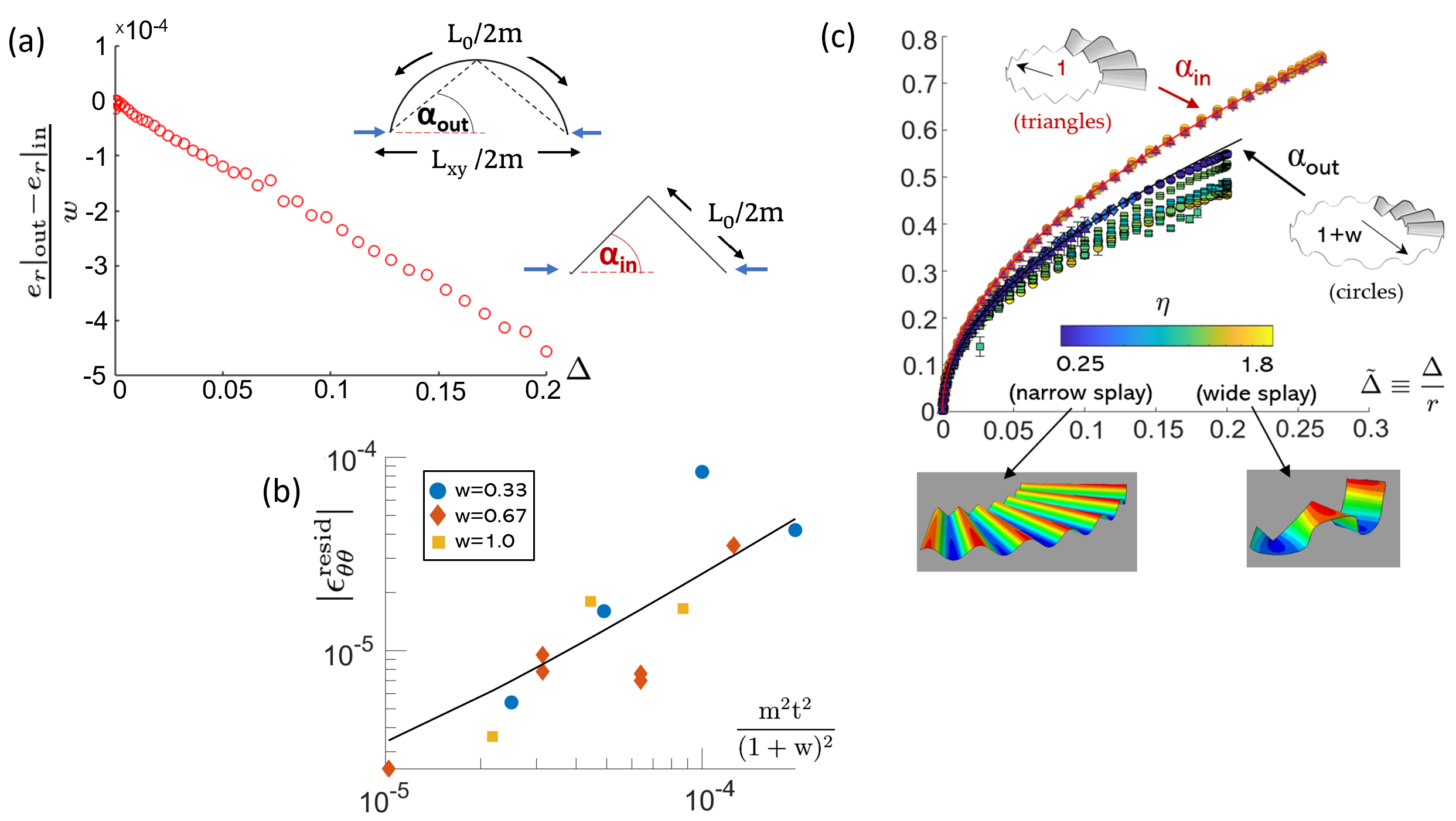}
\caption{Comparisons of geometric quantities between theory and simulations (using the explicitly periodic initial conditions described at the beginning of Sec.~\ref{sec:results}). 
(a) Testing radial inextensibility. The graph shows an average radial strain inferred from Eq.~\ref{eq:inextensibility}, i.e.~the small relative difference between the inward displacement at the outer boundary and that at the inner boundary, further discussed in Sec.\ref{sec:radialStrain}  This example had $w =0.33$, $m=25$, and $t = 2.67 \times 10^{-4}$.  (inset) Sketches of the wrinkle profile at
the two boundaries, showing the definition of lengths $L_0$ and $L_{xy}$, and slope angles $\alpha_{\rm out}$ and $\alpha_{\rm in}$. The blue arrows indicate the compression direction.
(b) Testing azimuthal inextensibility at the outer boundary. The magnitude of the residual azimuthal strain $|\epsilon_{\theta\theta}^{\rm resid}|$ recorded slightly inside $r=1+w$, plotted logarithmically against the collective variable $m^2 t^2/(1+w)^2$ at large $\Delta$, for multiple values of $m$, $t$ and $w$. The black line shows the linear fit: $|\epsilon_{\theta\theta}^{\rm resid}| \approx 0.24 \frac{m^2 t^2}{(1+w)^2} $.
(c) Wrinkle slope angle vs.~applied contraction $\tilde \Delta \equiv \Delta/r$ at the inner and outer boundaries (see Sec.~\ref{sec:results_slopeComparison}). (Top row) We separately test Eq.~(\ref{eq:alphaIn}) for $\alpha_{\rm in}$ (in red) and Eq.~(\ref{eq:alphaOut}) for $\alpha_{\rm out}$ (in black). The simulation data points are plotted alongside the model predictions. Simulated samples of differing width $w$ and wavenumber $m$ are coloured using the dimensionless cone splay parameter, $\eta \equiv \frac{\pi (1+w)}{m w}$. The error bars are standard deviations from averaging over all the wrinkles. Markers: $w=0.33$ (circles), $w=0.67$ (squares), $w=1.0$ (diamonds), $w=1.67$ (triangles). (Bottom row) Example wrinkle profiles for small and large $\eta$: (left) for $\eta = 0.25$ ($w=1.0$, $m=25$), and (right) for $\eta = 1.8$ ($w=0.33$, $m=7$). Colouring denotes height.
}
\label{fig:4}
\end{figure*}

First, we consider \textit{radial} inextensibility. Here, we  use approximations appropriate for fine buckling and narrow wrinkles as in the far-from-threshold regime of Ref.~\cite{Davidovitch2011}. A material circle at radius $r$ in the annulus has fixed initial arc length $L_0(r) = 2\pi r$. After the contraction, this circle deforms into a wavy line, whose horizontal projection is approximately circular under the observed fine buckling.  The radius of this contracted circle is the original $r$ plus some radial displacement $e_r$ due to the inward contraction. 
The fractional change in the projected length $L_{xy}$ is thus given by
\begin{equation}\label{eq:inextensibility}
L_{xy}/L_0 = 1 + e_r/r.
\end{equation}
At the inner boundary ($r=1$), $e_r = -\Delta$ because of the imposed boundary conditions. However, at the outer boundary ($r=1+w$), $e_r$ is free to assume any value. If it's different from $-\Delta$, this implies the presence of radial strain. In Fig \ref{fig:4}a, we use Eq. \ref{eq:inextensibility} to provide a simple estimate of the radial strain.  For the outer-boundary displacement we infer $e_r$ from Eq. \ref{eq:inextensibility} and subtract the inner boundary displacement $\Delta$.  This gives the change in the horizontal component of the radial length.  We interpret this small negative apparent strain in Sec. \ref{sec:radialStrain}.   
\par
We may gauge \textit{azimuthal} extensibility by measuring the full (not projected) change of arc length for the circle at radius $r$. Here we consider only the outer boundary region. As reported in Fig \ref{fig:4}b, there is a nonzero compression of these circles. This compression reflects the {\em Elastica} bending stress of pure bending mentioned in Sec.~\ref{sec:innerLame}. Specifically, bending at a scale $\lambda$ produces a residual bending stress, $\sigma_{\theta\theta}^{\rm resid} \sim -B/\lambda^2$, where $B$ is the bending modulus. Thus, from wrinkling of wavenumber $m$, we expect the outer boundary to have an azimuthal bending \textit{strain} given by:
\begin{equation}\label{eq:bendingStrain}
  \epsilon_{\theta\theta}^{\rm resid} \sim -\frac{1}{Y} \frac{B}{\lambda^2} \sim - \frac{m^2t^2}{(1+w)^2},
\end{equation}
where we have used $\lambda = 2\pi(1+w)/m$ and $B/Y \sim t^2$ ($t$ is the thickness). The strain $\epsilon_{\theta\theta}^{\rm resid}$ is calculated for our numerical samples by directly measuring the change in azimuthal bond lengths along a material circle slightly inside $r=1+w$ \footnote{This is done to avoid possible strain originating in Gaussian curvature boundary layers at the free outer boundary of the annulus; see \cite{Efrati2009}.}, and then averaging. $\epsilon_{\theta\theta}^{\rm resid}$ is negative as expected; moreover, it is roughly independent of $\Delta$ beyond a transient initial phase. Fig.~\ref{fig:4}b shows a log-log plot of $|\epsilon_{\theta\theta}^{\rm resid}|$ for multiple samples with differing values of $m$, $t$ and $w$. We see data collapse over the range of a decade, with the predicted slope of unity.We attribute the large variability of the points to variations in the specific wrinkle pattern, which change from one simulation to the next. Based on the linear fit, the estimated pre-factor for Eq.~\ref{eq:bendingStrain} is 0.24. We note that this is roughly $50$ times greater than the pre-factor calculated theoretically for purely rectangular \textit{Elastica} bending, which gives $\sigma^{\rm resid} = -2B/\lambda^2$ \cite{Davidovitch2019}. Such a larger value, however, is not unexpected given the inherently two-dimensional nature of our conical \textit{Elastica} arcs.
\par
Figs.~\ref{fig:4}a and \ref{fig:4}b together provide suggestive evidence for radial and azimuthal inextensibility of the numerical solutions. 

\subsection{Evolution of wrinkle amplitude} \label{sec:results_slopeComparison}
We now compare specific features of the wrinkle shape to the predictions of the cone-triangle construction. In Fig.~\ref{fig:4}c, we consider the amplitude of the observed wrinkles. A dimensionless way of measuring the amplitude is by determining the average slope angle $\alpha$ of the wrinkles at the inner and outer boundaries. Though the cone-triangle construction can be implemented for a wide range of wavenumbers $m$ for any given $\Delta$ and $w$, we considered only the narrow range of $m$ that appeared spontaneously in the unperturbed numerics, where $7 \lessapprox m \lessapprox 35$ \cite{Pal2023}.
\par
We define the average slope angle $\alpha$ to be the slope angle of the line joining the zero-crossing of the wrinkle with the adjacent maximum (illustrated in inset of Fig.~\ref{fig:4}a). For angle $\alpha_{\rm in}$ at the inner boundary, the figure explicitly relates $\alpha_{\rm in}$ to $L_{xy}/L$. For fine wrinkles, we may use Eq.~\ref{eq:inextensibility} to get: 
\begin{equation}\label{eq:alphaIn}
\cos \alpha_{\rm in} = 1-\Delta \equiv 1-\tilde \Delta(1), 
\end{equation} 
where we have introduced the reduced variable $\tilde \Delta (r) \equiv \Delta/r$.
At the outer edge, $L_{xy}/L$ is likewise the ratio of the horizontal line to the arc length of the circular sector (see inset of Fig.~\ref{fig:4}a). For our circular cone model, it is related to the slope angle $\alpha_{\rm out}$ by 
\begin{equation}\label{eq:alphaOut}
\sin 2\alpha_{\rm out}/(2\alpha_{\rm out}) = L_{xy}/L \equiv 1-\tilde \Delta(1+w).
\end{equation}
In Fig.~\ref{fig:4}c, we compare these cone-triangle slope angles to the simulated ones as a function of $\tilde \Delta(r)$ for both inner ($r=1$) and outer ($r=1+w$) boundaries. The upper red curve and data points compare $\alpha_{\rm in}$. The agreement of the data points with the curve confirms that each triangle segment tilts to preserve its length as it is drawn towards the axis.
The solid black curve gives the cone-triangle prediction of Eq.~\ref{eq:alphaOut} for $\alpha_{\rm out}$. The data points on and beneath the curve represent annuli of different widths $w$ and wavenumber $m$, coloured by their splay: $\eta \equiv \pi (1+w)/(m w)$. In general, our circular cone model is consistent with the prediction for narrowly-splayed wrinkles, but over-estimates the slopes for widely-splayed ones. The pictures below the graph show the annulus shape for different amounts of splay $\eta$. They suggest the source of the over-estimated slopes: for the widely splayed example with the greatest discrepancy, the boundary curve is significantly flatter than a circular arc. Thus the simplifying assumptions of our circular-cone construction are not satisfied in this widely-splayed regime.

\subsection{Comparing energies}\label{sec:results_energyComparison}
Perhaps the main significance of this isometric deformation is that its energy behaves so differently from that of a generically deformed sheet. It is different in two respects. Firstly, the energy depends only on the bending modulus $B$, and is independent of the stretching modulus $Y$ for given bending modulus. Secondly, the energy in this limit is qualitatively smaller than for a generic wrinkling deformation with curvature and thickness comparable to our simulations. Below we verify that the energy of the contracted annulus has these distinctive features. 

The elastic energy is a sum of strain energy and bending energy: 
\begin{equation*}
 U_{\rm elastic} = U_{\rm strain} + U_{\rm bend}
\end{equation*}
Thus, for a strain-free isometric configuration, we have $U_{\rm elastic} \approx U_{\rm bend}$. For an arbitrary conical deformation extending between lengths $L_{\rm min}$ and $L_{\rm max}$ along the cone axis, the bending energy is given by:
\begin{equation}\label{eq:Ubend_cone_kappa}
 U_{\rm bend}^{\rm cone} = \frac{B}{2} \log (\frac{L_{\rm max}}{L_{\rm min}}) \int ds \; \kappa_N^2,
\end{equation}
where $B$ is the bending modulus and $\kappa_N(s)$ is the curvature of the outermost arc of the cone normal to the surface, parameterised by the arc length $s$. For a circular cone as in our model, $\kappa_N(s) = \rm{const.}$. For fine wrinkles \footnote{i.e. for $\eta \ll 1 \leftrightarrow m \gg \rm{max}\{1,1/w\}$ },
 we can write $\kappa_N \approx 1/\rho_c$, where $\rho_c$ is the radius of curvature of the outer circle of the cone (see SI Fig.~\ref{fig:SI_fig2}b). Then, adding up the contributions of $2m$ cones for a wrinkled solution of wavenumber $m$ (see SI for details), we get:
\begin{equation}\label{eq:Ubend_theory}
 U_{\rm elastic}^{\rm model} (\Delta) = U_{\rm bend}^{\rm model} (\Delta) \approx B \log (\frac{L_{\rm max}}{L_{\rm min}}) \frac{ \pi (1+w) w}{\rho_c^2(\Delta)}. 
\end{equation}
Also, in order to avoid high-curvature regions near the inner boundary, where the thin-sheet approximation $\kappa t \ll 1$ becomes questionable, we choose to compare energies for only the outer $3/4^{th}$ part of the annulus, so that $\log (\frac{L_{\rm max}}{L_{\rm min}}) = \log(4)$.

In Eq.~(\ref{eq:Ubend_theory}) only the quantity $\rho_c(\Delta)$ is not explicitly known. Our circular cone construction (see SI) permits us to get an expression for $\rho_c (\Delta)$ in terms of the outer slope angle $\alpha_{\rm out}$:
\begin{equation}\label{eq:rhoc}
 \rho_c(\Delta) \approx \frac{\pi (1+w)}{4m} \frac{1}{\alpha_{\rm out}(\Delta)} .
\end{equation}
We recall that $\alpha_{\rm out}$ depends only on $w$ and $\Delta$, and not on $m$. Thus $\rho_c \sim 1/m$, and so $U_{\rm elastic}^{\rm model} \sim m^2$, which is the quadratic scaling with wavenumber (at fixed amplitude) expected for a bending energy. 
\begin{figure}[htb]\centering
\includegraphics[width=0.45\textwidth]{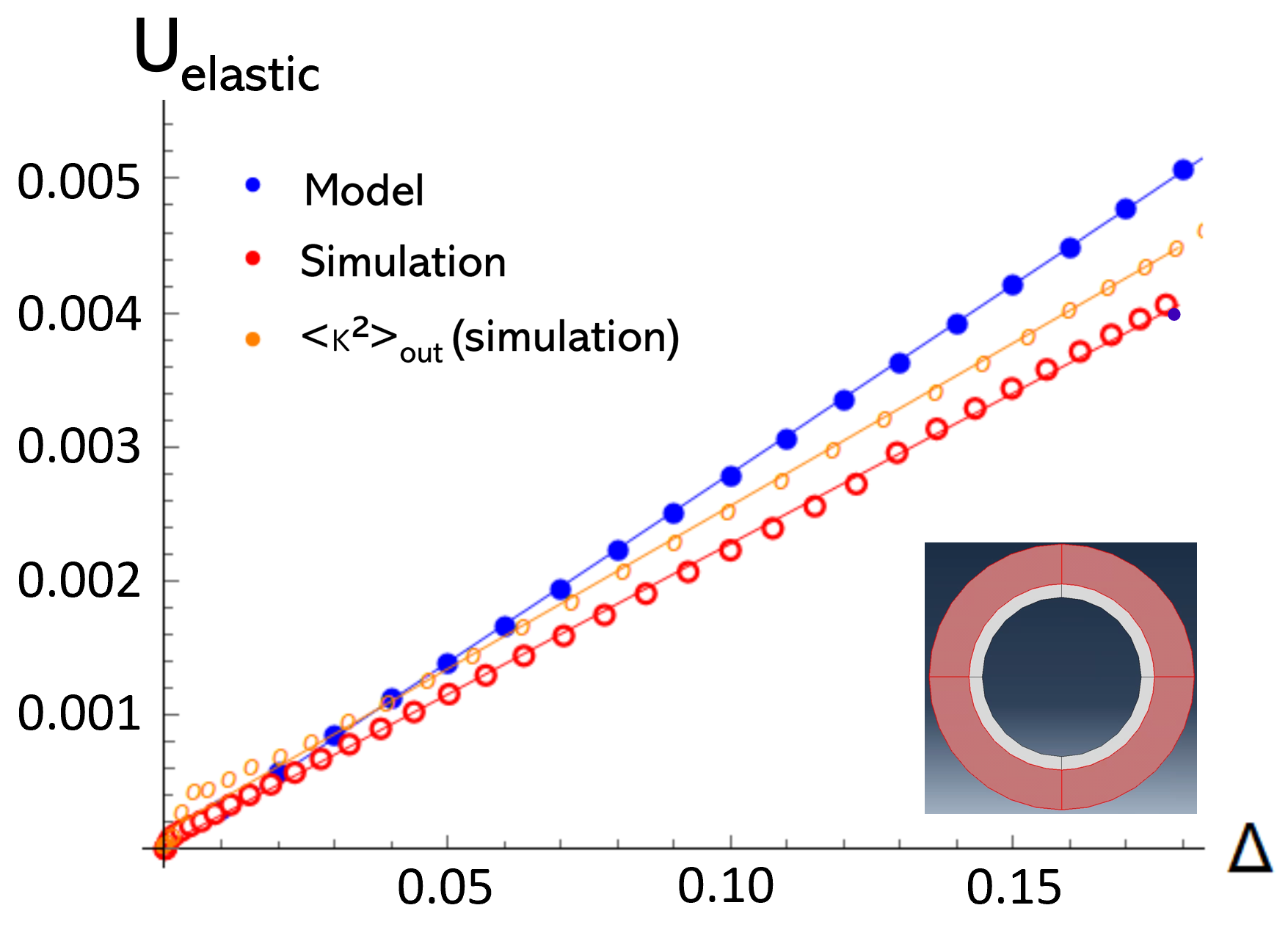}
\caption{Comparing elastic energies (see Sec.~\ref{sec:results_energyComparison}) for a representative sample ($w=1.0$, $m=25$, $t=6.67\times 10^{-4}$): $ U_{\rm elastic}^{\rm sim} (\Delta)$ ( lower curve) ; see \textit{Methods}) with $U_{\rm elastic}^{\rm model} (\Delta)$ ( upper curve); Eq.~\ref{eq:Ubend_theory}) and $U_{\rm \langle \kappa^2\rangle}^{\rm sim} (\Delta)$ ( middle curve); Eq.~\ref{eq:Ubend_cone_kappa}). All three curves show \textit{Elastica}-like linear scaling (solid lines are fits), but the two theoretical curves overestimate the simulated energy by roughly $15\%$ and $30\%$ respectively. Other $w$ and $m$ values also generally over-estimated the energies, as discussed in Sec.~\ref{sec:results_energyComparison}. Energy units are defined in \textit{Methods}. (inset) Annulus, indicating the region where energies were compared (in red), omitting the high-curvature regions near $r=1$.}
\label{fig:5}
\end{figure}

Our primary subject of interest, however, is the scaling of $U_{\rm elastic}^{\rm model}$ with $\Delta$ for fixed $m$. Since $\alpha_{\rm out}(\Delta) \sim \sqrt{\Delta}$ (see Fig.~\ref{fig:4}c), we expect the squared curvature and the bending energy to scale linearly with $\Delta$: $U_{\rm elastic}^{\rm model} \sim 1/\rho_c^2 \sim \Delta$. We note that this linear form is the same as for the \textit{Elastica} energy of a rod under weak Euler buckling \cite{Audoly2010}, which is expected since each transverse arc of the cone undergoes Euler buckling with this same $\Delta$ dependence. 

 We may compare this isometric energy with that of conventional wrinkles \cite{Davidovitch2019} whose height $h(r, \theta)$ has the separable form $h = f(r) \sin(m \theta)$. As discussed in the SI, radial wrinkles of this form have substantial strain $\epsilon_c$ of order of 
$\epsilon_c \simeq \Delta/m^2$, much larger than the bending strain of Eq. \ref{eq:bendingStrain}, viz. $\simeq ~m^2 t^2$. Accordingly, for fixed $\Delta$ and $m$, the conventional strain $\epsilon_c$ and its associated elastic energy are much larger than their bending counterparts as $t \to 0$. The SI shows that $\epsilon_c$ is also large for the observed $m$ values: $\epsilon_c \approx 50 \epsilon_B$.

In Fig.~\ref{fig:5}, we plot three energy curves for a typical sample. These curves compare the simulated energy for a typical example, $U_{\rm elastic}^{\rm sim} (\Delta)$ ( lower curve), with two theoretical estimates: $U_{\rm elastic}^{\rm model} (\Delta)$ ( upper curve; see Eq.~\ref{eq:Ubend_theory}) calculated from our cone-triangle model, and $U_{\rm \langle \kappa^2\rangle}^{\rm sim} (\Delta)$ ( middle curve) calculated indirectly from the simulated solution through measurement of the mean squared curvature at the outer boundary: $\langle \kappa^2 \rangle_{\rm out} \equiv \frac{1}{2 \pi (1+w)} \int ds \; \kappa^2$ \footnote{Here, $\kappa$ is the arc curvature, without reference to the surface. As before, for fine wrinkles, we have $\kappa \approx \kappa_N$.}, and by using Eq.~\ref{eq:Ubend_cone_kappa}.
 The three curves are similar, differing by a roughly constant factor of order unity. Notably, the full simulated energy including any strain energy, is smaller than the model bending energies. This discrepancy is natural for the circular cone model, since this circular ansatz does not minimise the bending energy. The assumed uniform, circular curvature does not minimize the energy of the cone shape needed in order to bridge between the adjacent triangles. We expect that the system chooses a curvature profile $\kappa(s)$ that minimizes this energy. Thus, as expected, the use of the measured $\kappa(s)$ improves the agreement with the measured energy greatly. But a clear discrepancy remains.
 
We discuss the possible source of this discrepancy in SI Sec.~\ref{sec:departures}. It compares the simulated energy profile with that of a conical shape. The mismatch between the two profiles implies that the real surface does not have the straight director lines needed for a developably isometric surface. Such a mismatch requires strain energy. Accordingly, the mismatch should lessen with decreasing thickness. On the other hand this mismatch enables a reduction in bending energy below that of a cone. Thus the SI suggests how a small strain distortion can reduce the total energy below our isometric estimate, as the figure shows.


We conclude that the elastic energy of deformation for the wrinkles encountered in this study is strongly consistent with our claim of an isometric conical deformation. This minimal geometric picture gives a quantitative account of the structure and energy for a given wave number $m$. Residual discrepancies seem consistent with expected residual stretching effects. Furthermore, the explicit circular-cone-triangle construction gives a workable quantitative estimate of the full elastic energy. This approximation is least accurate when the splay parameter $\eta$ is large. 

\section{Discussion}\label{sec:discussion}
\subsection{A developable wrinkling morphology}
As shown above, the Lam\'e radial contraction system is qualitatively altered by removing the outer tension from that system. Specifically, the purely geometric definition of our deformation results in a faceted shape that is in clear contrast to prior force-based deformations \cite{Davidovitch2011, Huang2007,pineirua_capillary_2013}. Our work demonstrates a strong resemblance between the observed shape and the exactly isometric cone-triangle shape defined above. For the numerically feasible thicknesses we report, the cone-triangle hypothesis allows us to reproduce structural and energetic features to good accuracy. Still, detailed analysis shows clear-cut departures from this shape that remain unexplained. 

Developable approximations such as the cone-triangle model appear to be a novel approach to describing periodic buckled structures. For the annular contraction deformation studied here, this developable picture seems entirely adequate as an initial quantitative account of the shape. This contrasts with other faceted shapes\cite{Witten2007,Seffen2014}such as twisted ribbons ~\cite{PhamDinh2016}, where a comparable understanding of the energy requires explicit consideration of non-developable geometry and non-bending energy. Our cone-triangle decomposition also provides an instructive example of a buckled shape in a radially symmetric system that does not have the conventional separable form -- i.e.~the buckling displacement is not a product of a radial factor and azimuthal factor.

The developability of the faceted wrinkling structure makes it fundamentally geometric, so that it can be created easily on a computer. This provides a way to programmatically construct \textit{smooth} radially wrinkled patterns in materials using intrinsically flat components, much in the spirit of origami and other designed materials \cite{Callens2018}. In fact, a well-known real-world example of such programmed radial wrinkling is the Elizabethan ruff collar. One way of making a ruff is exactly using the inner Lam\'e boundary conditions: we take a cloth annulus, and manually make pleats in it such that the inner boundary of the annulus lives on a contracted radius. Figs.~\ref{fig:1}e and \ref{fig:1}d show, respectively, a ruff collar and its reconstruction using our circular-cone-triangle model; the visual similarity between the two is striking. The conical outer solution seems to capture the ruff wrinkles, in particular their radial straightness and their shape at the outer boundary. 

 As noted above, accounting for the wrinkles of Fig.~\ref{fig:1}c is not equivalent to finding a ground-state configuration consistent with the imposed boundary condition. Our deformation was performed quasi-statically, and the wrinkles appeared continuously without abrupt jumps. Thus the system followed a single energy minimum as it deformed.  This observed deformation may well have significantly higher energy than the ground state has. 

\subsection{Nature of the thin-sheet limit}
The wrinkles investigated here closely approximate the isometric cone-triangle structure, in both shape and energy, as shown above. That is, many numerical examples with a broad range of thicknesses are well described by a model with no dependence on thickness. We take this as strong evidence that, for a given fine wrinkle number $m \gg \rm{max} \{1,1/w\}$, the sheet approaches this isometric shape as thickness goes to zero. This leaves open the question of how the system selects this $m$, which clearly {\em does} depend on thickness. Given the present understanding of the energy and shape, the groundwork is laid to explain how this wrinkle number is selected. A separate paper by one of us \cite{Pal2023} addresses this question. 

\subsection{ Approximations and limitations}
\subsubsection{Circular cone {\em Ansatz}}
The key modelling assumption used in this paper is the circular {\em Ansatz} for the bent cones. As we have shown, this {\em Ansatz} serves as a serviceable approximation to the true energy-minimising \textit{Elastica} solution, while being simple enough to allow for explicit calculation of several key observables, namely, the slope $\alpha_{\rm out}$ and the energy $U_{\rm elastic}$. Sec.~\ref{sec:results} shows that this approximation is accurate only for relatively narrow cones (small $\eta$) and large $w$. This can be partially explained as follows. We expect the circular {\em Ansatz} to hold true only for \textit{weak} Euler buckling. ``Weak" here means that the angular squeezing of the cones and the resultant slope, are both small. In the SI, we show that this squeezing experienced by a conical sector during deformation increases sharply with decreasing $w$, such that for our narrowest annuli, it can be as high as $20\%$ for our maximum $\Delta$.
At such large values, one no longer expects the weak Euler regime to be valid, and the resultant shape manifests its non-linearity through the flattening apparent in Fig.~\ref{fig:4}c. We note, however, that even at such large contractions, both the circular {\em Ansatz} and the numerical simulations give a linear energy profile, although their slopes are no longer close.
\subsubsection{Radial Strain} \label{sec:radialStrain}
Our main quantitative evidence for isometric wrinkling comes from the azimuthal strain of Fig.~\ref{fig:4}b and the total energy of Fig.~\ref{fig:5}. We were not able to obtain comparable quantitative evidence by measuring the radial strain.  The nominal radial strain reported in Fig.~\ref{fig:4}a is of the same order as the azimuthal strain, but it is negative, contrary to expectations.  
These small strains are significantly affected by further effects that we did not include.  One such effect is the vertical slope of the material radial lines.  For example a material line extending radially from the tip of an inner zigzag to the peak of an outer wrinkle undergoes a change of height $z$ from $z_{\rm in}$ to $z_{\rm out}$. Now, the estimate of Fig.~\ref{fig:4}a amounts to approximating this line by its horizontal projection, $\ell_{xy}$. The full displacement between inner and outer ends of the line is thus $\sqrt{\ell_{xy}^2 + (z_{\rm in}-z_{\rm in})^2}$. For the data point in Fig.~\ref{fig:4}a at $\tilde \Delta = 0.2$, this effect adds a correction to the strain of +.0002 to +.0003, greatly reducing the reported value of strain. In addition,  the material line is necessarily longer than displacement between its inner and outer ends. Further corrections come from localised strain effects near the inner boundary, as discussed in the next section.  
\par
Because of such corrections Fig.~\ref{fig:4}a shows only a rough consistency with the asymptotic isometry we claim here. For stronger evidence we rely on the energy measurements of Fig.~\ref{fig:5}.

\subsubsection{Inner boundary approximation}
\label{sec:innerBoundary}
 Despite this strong resemblance between the simulated wrinkles and the cone-triangle structure, the match cannot be exact. Our construction of the cone-triangle shape in Sec \ref{sec:constructionSolution} required that the inner boundary be a polygon. It is only a good approximation to the simulated circular annulus when the wavelength ({\em i.e.} $2\pi/m$) is sufficiently small. In particular it must be small in relation to the width $w$ if the triangles are to resemble those of Fig.~\ref{fig:1}. Details are provided in the SI, Sec.~\ref{sec:SI_construction_limits}.

One clear disagreement concerns the length of the inner boundary. In the simulation, the undeformed inner boundary is a circle. Under contraction, each tilted circular arc segment must stretch if it is to remain on the surface of a cylinder. Indeed, we observed such stretching along the inner boundary nodes. However, this stretching zone was confined to this first ring of nodes, with a much weaker stretching in the adjacent ring. In addition, we verified that the associated elastic energy was much smaller than that reported in Fig.~\ref{fig:5}.

The inexactness of the triangle construction at the inner surface makes it clear that the construction is not a realistic representation of a circular annulus contracted onto a cylinder for \textit{arbitrary} wavenumber $m$. We demonstrated the isometric shape for only a limited range of geometries; both the displacement $\Delta$ and the width $w$ were confined to moderate ranges. For these moderate conditions the system appears to choose wavelengths fine enough that the inaccuracies of the cone-triangle construction are not important. For more extreme ranges, we expect that the cone-triangle picture shown here will need modification and that the $t \to 0$ limit will be more subtle. In particular, the locally stable isometric structure observed here may give way to macroscopic folding \cite{brau_wrinkle_2013}.
Finding the conditions in which the cone-triangle wrinkles occur is a fruitful subject for future work. 

\subsubsection{Lateral constraint boundary condition}
\label{sec:lateralConstraint} 
As noted in \textit{Methods}, we prohibit any lateral movement of the nodes on the inner boundary, in order to stabilise the faceted wrinkled regime against folding instabilities. However, even if we do allow lateral movement, we see the same faceted wrinkling at early simulation times, before azimuthal movement of the boundary eventually leads to folding\cite{brau_wrinkle_2013, Paulsen2017}. That is, the, excess material concentrates at a single sector along the circumference. Our observations suggest that faceting can be seen for initial buckling, even in the absence of lateral constraints. But our evidence is still only suggestive and needs to be confirmed through physical experiments. 

\subsection{Application}
The deformation geometry treated here amounts to a strategy for avoiding the typical fate of deformed sheets. Such sheets adopt structures that require significant elastic energy beyond the needed bending energy. Often these structures produce uncontrolled disorder, {\em e.g.\ } as seen in a crumpled sheet. By contrast, the radial contraction used here guides the sheet into a locally low-energy channel resulting in only bending energy. This low-energy state is moreover attainable by continuous deformation from a flat reference state. Since these deformations lack the additional deformation costs of competing structures, they are intrinsically (locally) stable. They provide a general means to fold and compact a sheet continuously and reversibly with minimal external guidance. We note that the curved inner boundary appears important in stabilising this structure; straight boundaries with no radial contraction \cite{Vandeparre2011} do not produce analogous regular wrinkling.

These guiding channels should appear more generally than in the simple circular geometry studied here. The same principles control other inwardly curved boundaries, and other directions of contraction, e.g.~onto a constraining cone instead of a cylinder.

\subsection{Conclusion}
While the emergence of developability in wrinkled systems has been studied intensively over the past decade, the simplicity and symmetry of the inner Lam\'e geometry makes it a particularly clean realisation of such a phenomenon. This faceted morphology is qualitatively different from previously known forms of two-dimensional wrinkling. Significantly, it creates sharply dictated, strongly distorted features with negligible stretching and without a separable radial/azimuthal shape. Moreover, it has many variants to be explored, with real world applications. Finally, this work sets the stage to explain the selection of the wavenumber $m$ \cite{Pal2023}.

\section{Materials and methods}\label{sec:methods}
To investigate the wrinkling morphology of the inner Lam\'e system, we used the commercial finite-element (FE) solver Abaqus 2018 (SIMULIA, Dassault Systèmes). Using both implicit and explicit dynamic methods \footnote{`Dynamic' means through time integration of the equations of motion. `Implicit' (using a modified Newton-Raphson method) and `explicit' refer to the method of integration.} in a quasi-static regime, we performed simulations of the inner Lam\'e system using its defining boundary conditions: radial displacement $e_r(r=1)=-\Delta$ at the inner boundary mesh points, with the outer boundary nodes (at $r=1+w$) kept free. The maximum value of $\Delta$ applied was $\Delta_{\rm max}=0.267$. 

The simulation seeks the configuration that would be attained by adiabatically contracting the inner boundary (i.e., its nodes). Accordingly, it  gradually decreases $e_r$ from zero to $-\Delta$ as time $t$ increases from $0$ to $t_f$.  The value of $t_f$ is made sufficiently large that the kinetic energy is always a small fraction of the potential energy.  Then the final configuration was not noticeably affected by further increase of $t_f$, as detailed in the SI.

To account for the possibility of high strains, the sheet was initially modelled as a Neo-Hookean hyperelastic material with coefficients equivalent to the linear moduli: Young's modulus, $E=0.907125$ MPa, and Poisson ratio, $\nu =0.475$ (c.f.~Eq.~\ref{eq:Ubend_theory}), corresponding to a rubber-like material. The associated bending stiffness $B$ with thickness $t$ is then given by $B= \frac{E t^3}{12 (1-\nu^2)}$. To verify that results are independent of the material model, we also re-performed several simulations with a linear material model with these same moduli. Further details on our FEA methods are given in the SI.

To test the validity of our results over a range of parameters, we kept the inner radius fixed and varied the other two parameters -- width $w$ and thickness $t$ -- over the range of a decade. The width $w$ of the annulus was varied between $w=0.33$ (very narrow) to $w=1.67$ (moderately wide) -- a factor of almost 10. The thickness $t$ was varied between $t=2.67\times 10^{-3}$ (thick) to $t=1.33\times 10^{-4} $ (very thin), a factor of 20. We performed consistency checks to ensure that the final morphology was independent of the choice of any simulation parameters. For data extraction, we used the software package Abaqus2Matlab \cite{Papazafeiropoulos2017}. 

We used two principal deformation protocols to ensure that the wrinkles we observed were characteristic of the quasi-static limit. Our baseline protocol was done using explicit dynamics, in which the finite elements are accelerated with a constant damping factor. The other deformation protocol used an implicit scheme for integrating the forces, with the damping factor automatically selected by the software to favour quasi-static behaviour. In both cases, to avoid gross kinetic effects, we increased $\Delta$ from 0 to $\Delta_{\rm max}$ at a rate slow enough such that the kinetic energy of the system always remained $\lessapprox 5$\% of the elastic energy. This is standard procedure for quasi-static analyses in finite-element simulations (see SI for details). 

The specifics of the wrinkling morphology depend on how the contracted boundary is constrained. For example, the clamped boundary studied by Mora and Boudaoud \cite{mora_buckling_2006} leads to a highly compressed boundary layer whose relation to the sheet is difficult to discern. The other extreme is to allow each point on the boundary circle to lie anywhere on the confining cylinder; such boundaries lead to wrinkling only in a transient regime, leading ultimately to collapse into a fold \cite{pocivavsek_stress_2008,brau_wrinkle_2013}. Here we study an intermediate constraint in which points on the bounding circle may move only axially (i.e.~vertically) on the bounding cylinder; azimuthal ($\theta$) motion is not allowed. Doing so automatically prohibits folding (since this requires lateral motion), but without interfering with the wrinkling. 
\section*{Author Contributions}
A.S.P. conceptualised the problem, performed simulations and measurements, and created the majority of the text and figures. L.P. was key in providing numerical support, and in critical readings of several drafts. T.A.W. supervised this work, formulated the cone-triangle model, and helped with conceptualisation, interpretation of data, writing and revising.

\section*{Conflicts of interest}
There are no conflicts to declare.

\section*{Acknowledgements}
The authors would like to thank Benny Davidovitch, Dominic Vella and Enrique Cerda for many insightful discussions, and Nhung Nguyen and George Papazafeiropoulos for help with numerical analysis. This work was partially funded by the National Science Foundation (NSF)-Materials Research Science and Engineering Center (MRSEC) grants DMR-1420709 and DMR-2011854.



\balance


\bibliography{rsc-articletemplate-softmatter} 
\bibliographystyle{rsc} 

\section{Supplementary Information}
\subsection{Construction of circular conical sections}\label{sec:SI1_coneConstruction}
In the main text, we describe how we divide the flat annulus into two regions -- an inner region composed of isosceles triangles, and an outer region composed of arc sectors. The main text describes how the inner triangles can isometrically deform under an inner displacement (or contraction) of length $\Delta$ by tilting about the normal to the inner boundary -- this gives an inner `buckled' solution. Here, we describe how to construct the outer solution whereby the arc sectors isometrically bend into sections of right-circular cones. In other words, each arc bends into a circular profile.
\begin{figure}[htb]
\centering
\includegraphics[width=0.5\textwidth]{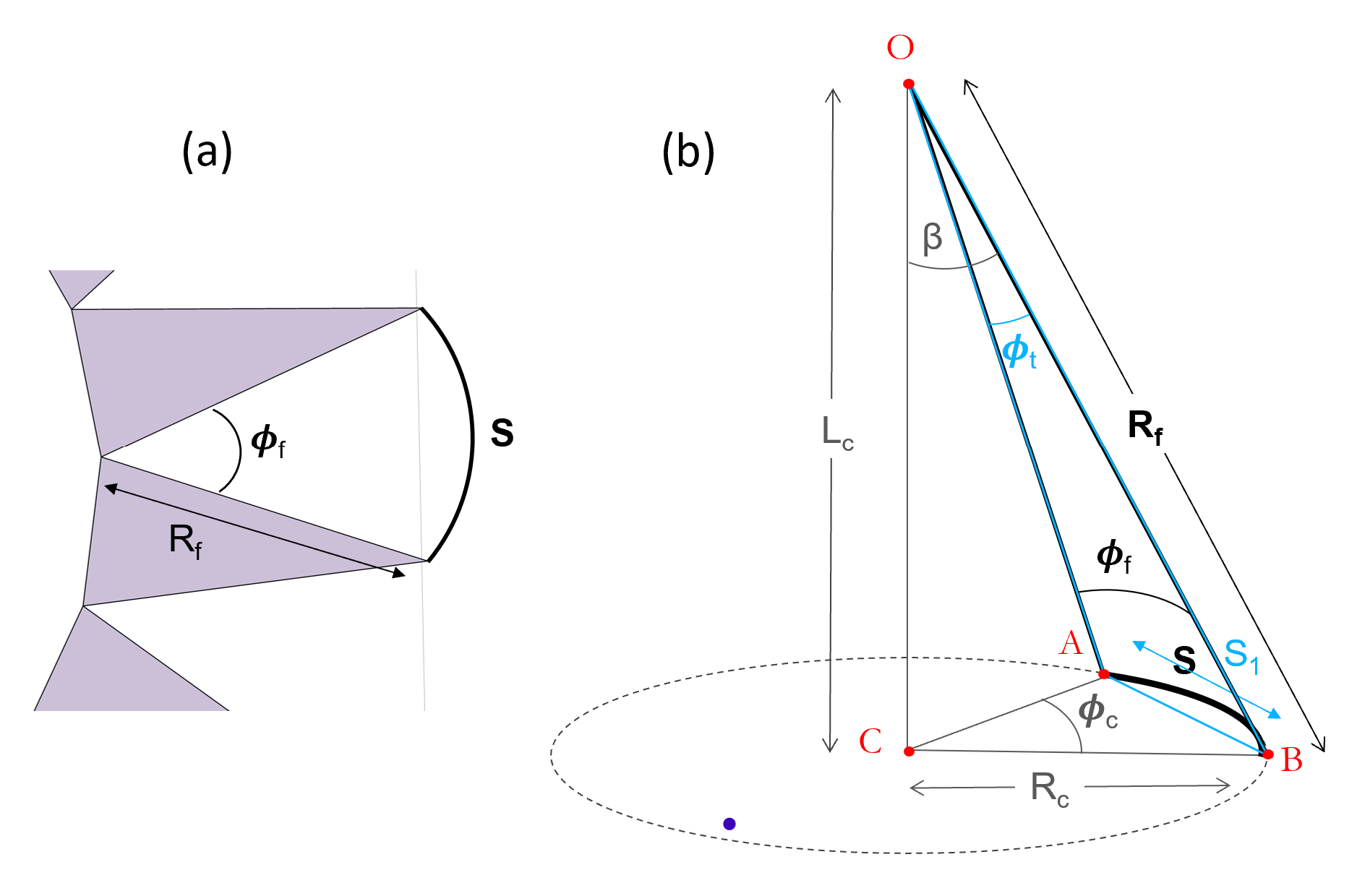}
\caption{Isometrically constructing a sector of a cone from a flat arc sector. (a) The flat configuration.~(b) The curved configuration; the actual conical sector and its variables are depicted in black, the triangle $\bigtriangleup AOB$ formed by the edges of the two adjacent tilted triangles, along with its variables, is in blue, and the putative cone with its variables is in grey.}
\label{fig:SI_fig1}
\end{figure}
\par
\begin{figure*}[htb]
\centering
\includegraphics[width=1.0\textwidth]{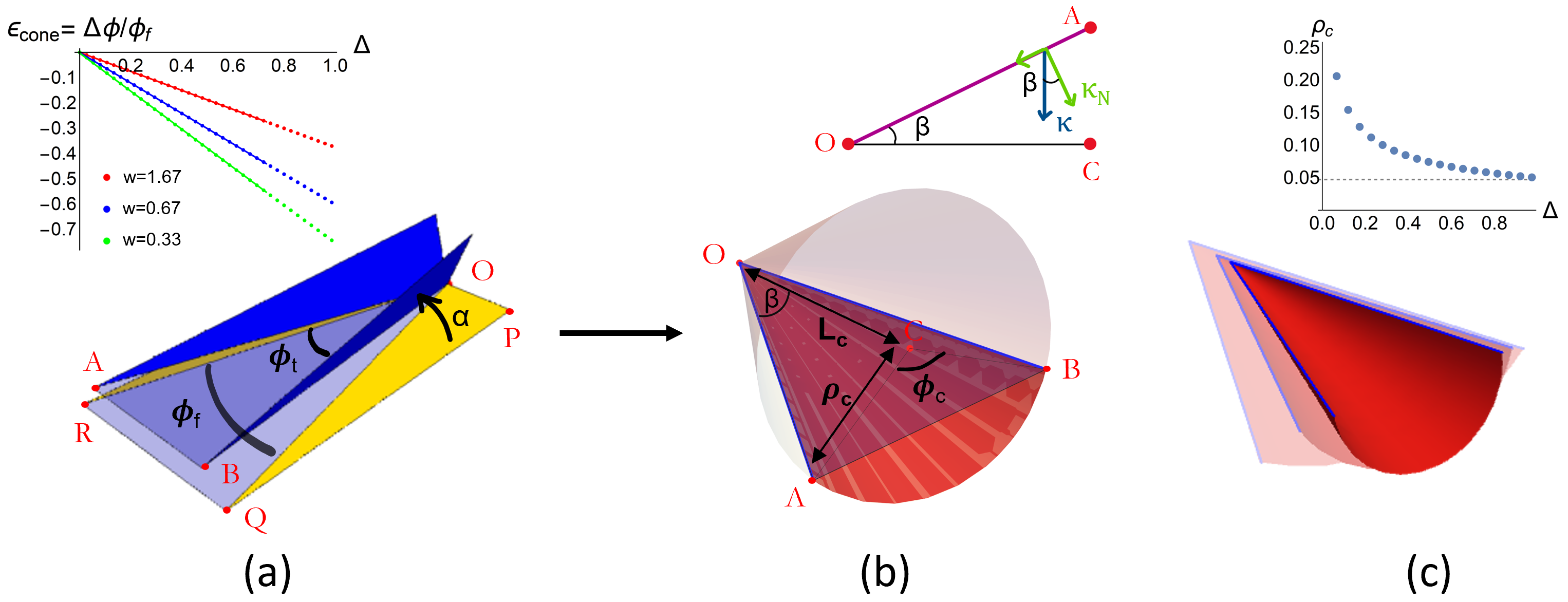}
\caption{Constructing a single cone 
(a) Two adjacent flat triangles (in yellow) subtending an angle $\phi_f$ tilt towards each other by angle $\alpha$. The tilted triangles (in deep blue) subtend a smaller angle $\phi_t$. Thus, the region between the triangles (shown approximately as a violet triangle) gets \textit{squeezed} by an angle $\Delta \phi = \phi_t - \phi_f $. (inset) Evolution of the conical strain $\epsilon_{\rm cone} = (\Delta \phi)/\phi_f$ with $\Delta$ for three different widths $w$. Model measurements are plotted alongside the predicted lines of slope $= -1 /(1+w)$ (see Eq.~\ref{eq:SI_epsilonCone}).
(b) To fit within the contracted boundary defined by $OA$ and $OB$, the region must bend into a circular cone of radius $\rho_c$, angular extent $\phi_c$, axial length $L_c$ and tilt angle $\beta$. The only ingredients needed for the construction are $\triangle AOB$ and the tilt angle $\alpha(\Delta)$. (inset) Cross-section of the cone (in magenta) through $\triangle OAC$, showing that the arc curvature $\kappa$ and the normal curvature $\kappa_N$ are related by $\kappa_N = \kappa \cos \beta$. 
(c) The evolution of the conical solution with $\Delta$ (here, $0.05 \leq \Delta \leq 0.6$). Greater opacity means larger $\Delta$; the cones are translated with respect to each other for viewing clarity. (inset) As $\Delta$ increases, the radius of curvature $\rho_c$ decreases, and for $\Delta \to 1$, attains an asymptotic value of $\rho_c \approx \frac{\pi w}{2 m}$ (dotted line). 
}
\label{fig:SI_fig2}
\end{figure*}

The requisite variables for our geometric construction are described in Fig.~\ref{fig:SI_fig1}. In the flat state, each arc sector, lying between two adjacent flat triangles, subtends an angle $\phi_f$ at the common vertex of these two triangles. If the edges of the triangles have length $R_f$, then the length of the largest arc is $S= R_f \phi_f$. On tilting, the edges of these same triangles approach each other, so that they subtend a reduced angle $\phi_t$ (see SI Fig.~\ref{fig:SI_fig2}a). This reduction in angle ($\Delta \phi = \phi_t - \phi_f \leq 0$) constitutes the effective compressive strain $\epsilon_{\rm cone} = \Delta \phi/ \phi_f$ for the arc sector. SI Fig.~\ref{fig:SI_fig2}a shows the evolution of $\epsilon_{\rm cone}$ with boundary contraction $\Delta$. We see clearly that narrower sheets feel a greater $\epsilon_{\rm cone}$ for the same $\Delta$.

\par
The excess material can be accommodated by bending the sector into a cone. The exact shape of this cone can be determined by solving the \textit{Elastica} equation with the appropriate boundary conditions. However, for simplicity, in this paper we choose to approximate the exact shape by a circular cone. The aim then is to find the circular cone that fits in between the two adjacent triangles while exactly accommodating the excess angle $\Delta \phi$. 
Fig.~\ref{fig:SI_fig1}b shows all the variables needed for this operation. The isometrically bent conical section has to fit inside the triangle $\bigtriangleup AOB$ (drawn in teal) defined by the apex $0$ and the tilted edges $OA$ and $OB$ (see also Fig.~\ref{fig:3} in main text). The conical section itself is drawn in bold black. Let the circular cone have angular extent $\phi_c$, and let its maximum radius be $\rho_c$, centred at point $C$. Then the conical shape can be determined from the following set of geometric constraints.

\par
The first constraint is that of inextensibility, i.e.~length conservation. This gives us:
\begin{equation}\label{eq:SI_coneConstruct_inextensibility}
 S = \rho_c \phi_c = R_f \phi_f.
\end{equation}
The second constraint involves specifying the end-to-end distance $S_1$ between points A and B. Thus, using the blue triangle $\bigtriangleup AOB$ and the grey triangle $\bigtriangleup ACB$ in Fig.~\ref{fig:SI_fig1}b, we have: 
\begin{equation}\label{eq:SI_coneConstruct_end2end}
 S_1 = 2 \rho_c \sin(\phi_c/2) = 2R_f \sin(\phi_t/2).
\end{equation}
Eqns.~(\ref{eq:SI_coneConstruct_inextensibility}) and (\ref{eq:SI_coneConstruct_end2end}) constitute a set of simultaneous equations for the cone variables $\rho_c$ and $\phi_c$, since $R_f$ and $\phi_f$ are fixed by the flat geometry, while $\phi_t$ is set by the tilt of the neighbouring angles. Finally, using basic trigonometry, we have for the cone's apex angle:
\begin{equation}\label{eq:SI_coneConstruct_beta}
 \sin \beta = \rho_c/R_f.
\end{equation}
In sum, we have three independent equations for three unknowns: $\phi_c$, $\rho_c$ and $\beta$ (all positive-definite), which fully specify the cone in space (including the centre $C$). SI Fig.~\ref{fig:SI_fig2}b shows a representative solution of such a constructed cone. SI Fig.~\ref{fig:SI_fig2}c shows the evolution of this construction with increasing $\Delta$.

\subsection{Correcting for scalloped arc sectors}
The flat arc sectors as defined above -- circular arcs with their origin at the inner boundary -- lead to the creation of a flat shape that is in fact a \textit{scalloped} annulus, whose outer circumference $2m R_f \phi_f$ is greater than the expected $2 \pi (1+w)$.

We can correct for this length discrepancy when constructing the bent cones, by changing $R_f \to \chi R_f$ in Eq.~(\ref{eq:SI_coneConstruct_inextensibility}) where $\chi = \frac{\pi (1+w)}{m R_f \phi_f} \leq 1$. Eq.~(\ref{eq:SI_coneConstruct_end2end}) remains unchanged since it specifies the end-to-end-distance, which is set by the neighbouring tilted triangles. We note that this factor $\chi$ is smallest for samples with small $w$ and $m$ (wide splay $\eta$), and becomes $\approx 1$ for large $w$ and $m$ (narrow splay).

\begin{figure}[htb]
\centering
\includegraphics[width=0.35\textwidth]{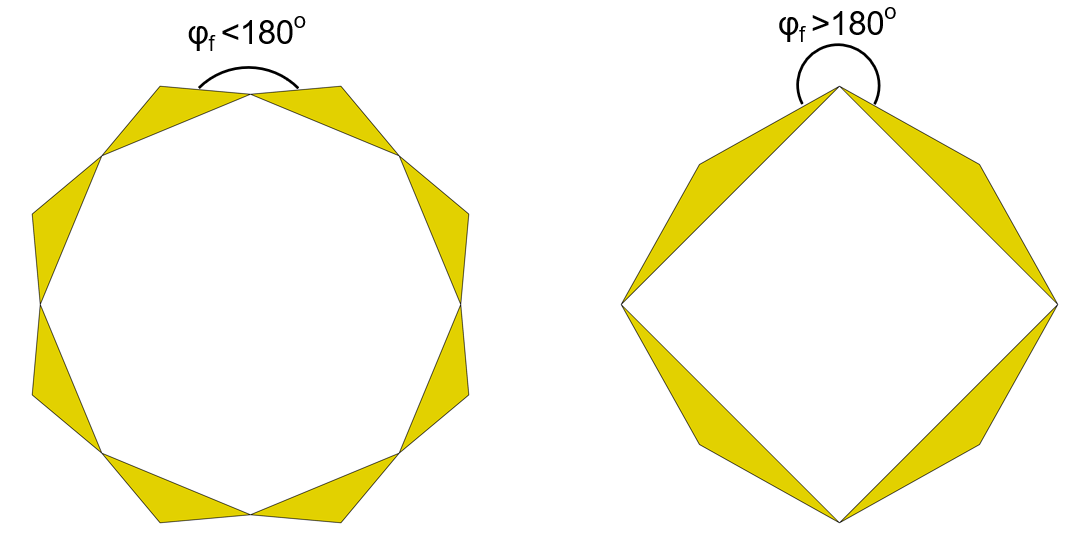}
\caption{Diagram supporting the calculation in Sec.~\ref{sec:SI_calculateConicalStrain}. Triangle $OPQ$ is the same as  in Fig.~\ref{fig:SI_fig2}a.  Here, we give the points explicit coordinates ( shown in black text). For more details, see main text.
}
\label{fig:SI_fig3}
\end{figure}

\subsection{Calculating the conical strain $\epsilon_{\rm cone} (\Delta)$}\label{sec:SI_calculateConicalStrain}
The inset of Fig.~\ref{fig:SI_fig2}a shows the evolution of the conical strain $\epsilon_{\rm cone}$ with contraction $\Delta$ for our cone-triangle construction. The observations (plotted with dots) show that $\epsilon_{\rm cone}(\Delta)$ is linear, with the slope depending significantly on the width $w$ of the sheet, but independent of the wavenumber $m$. To understand this behaviour, we analytically derive an expression for $\epsilon_{\rm cone} (\Delta)= (\phi_t - \phi_f)/\phi_f$ . 
\par
Since $\phi_f$ is fixed by the initial configuration, to find $\epsilon_{\rm cone}$, we only need to calculate $\phi_t (\Delta)$. The angle $\phi_t(\Delta)$ can be calculated given any one of the inner triangles, the flat arc angle $\phi_f$, and the tilt angle $\alpha$. We use the triangle $\triangle OPQ$ from Fig.~\ref{fig:SI_fig2}a, located in the $x-y$ plane. The annular width $w$ gives the height of the triangle, while $h$ gives its base.  Now, instead of rotating about $O$, we choose to rotate the triangle about its mid-line. For this, we let $X$ be the mid-point of $OP$, and we set it to be the origin $(0,0)$, as shown in 
Fig.~\ref{fig:SI_fig3}. In similar fashion,  we define $Z$ to be the midpoint of the edge $QR$, so that the bisector of the cone is the line $OZ$, shown as a dotted blue line in Fig.~\ref{fig:SI_fig3}. The angle between $OQ$ and the blue dotted line is then $\phi_f/2$. 
\par
This triangle is then tilted clockwise out-of-plane about $\overrightarrow {XQ}$ (parallel to the $y$-axis). Under this rotation, $P \to P'$ and $O \to O'$, while $X$ and $Q$ remain unchanged. In this tilted configuration, $\phi_t/2$ is given by the angle between the rotated vector $\overrightarrow {O'Q}$ and the vertical blue plane defined by $\overrightarrow{OZ}$. In what follows, we use vector notation to denote both 3d and 2d vectors, precising their nature where necessary.  The steps for determining $\phi_t$ are as follows:

\begin{enumerate}
\item Find the 3-vector $\overrightarrow{XO'} = \mathbb{R}_{-\alpha} \cdot \overrightarrow{XO}$, where $\mathbb{R}_\alpha$ is the standard 3x3 rotation matrix of angle $\alpha$ about the y-axis (positive $\alpha$ is considered anti-clockwise), and $\overrightarrow{XO} = (h/2,0,0)$ .
\item Then $\overrightarrow{O'Q} = \overrightarrow{XQ} - \overrightarrow{XO'} = (-h/2\cos\alpha, w, h/2\sin \alpha)$.
\item We only need its orthogonal projection on to the x-y plane: the 2-vector $\overrightarrow {O'Q_\perp} = (-h/2 \cos{\alpha}, w)$.
\item We need the vector angle between this and the vertical blue plane defined by $\overrightarrow{OZ}$. Here, $\overrightarrow{OZ} = \mathbb{R}_{-\phi_f/2} \cdot \overrightarrow{OQ}$, where $\overrightarrow{OQ} = (-h/2,w,0)$ .
\end{enumerate}

Finally, we have $\phi_t/2 =\mathrm{ VectorAngle[}\overrightarrow{O'Q_\perp}, \overrightarrow{OZ}\rm{]}$. Calculating this in a symbolic software like Mathematica, we get a complicated trigonometric expression that depends on three variables: $\alpha$, $\phi_f$, and the dimensionless triangle aspect ratio $h/w$. 
\par
To simplify this expression, we consider the regime of large $m$, which corresponds to both small $h/w$ and small $\phi_f$. Expanding to first order in $h/w$, and using the exact relation $\cos\alpha = 1-\Delta$ to replace $\alpha$ by $\Delta$, we get:
\begin{equation}
    \phi_t = \phi_f - \frac{h}{w} \Delta + \mathcal{O}((h/w)^2)
\end{equation}
Thus, we get for the cone contraction factor:
\begin{equation}
    \epsilon_{\rm cone} \equiv \frac{\phi_t-\phi_f}{\phi_f} \approx -\frac{h}{w \phi_f} \Delta.
\end{equation}
Substituting $h/w \approx \frac{\pi}{m w}$ and $\phi_f \approx \frac{\pi(1+w)}{m w}$ (valid for large $m$), we get:
\begin{equation}\label{eq:SI_epsilonCone}
    \epsilon_{\rm cone} \approx - \frac{1}{1+w} \Delta.
\end{equation}

\par
Eq.~\ref{eq:SI_epsilonCone} predicts that, for large $m$, the slope of $\epsilon_{\rm cone}$ vs.~$\Delta$ should be given by the ratio of inner to outer radius of the annulus. 
We plot this prediction as solid lines alongside the measurements in the inset of SI Fig.~\ref{fig:SI_fig2}a. The two are fully consistent. While Eq.~\ref{eq:SI_epsilonCone} for the slope has been shown only in the regime of narrow cones, in practice, we find it to be approximately valid even for wider cones. The qualitative conclusion here is that, for given $
\Delta$, wider annuli \textit{effectively} feel less squeezed than narrower annuli in a cone-triangle deformation, independent of the number of triangles/cones. 
\subsection{Calculating bending energy for a cone}
In this section, we continue with the variables introduced in Sec.~\ref{sec:SI1_coneConstruction}. A right-circular cone is made up of a series of circles whose radii $\rho$ increase linearly with distance (say, $\zeta$) along the cone axis, measured from the cone tip. For a cone of axial length $L_c$ (see SI Fig.~\ref{fig:SI_fig2}a) and maximum radius $\rho_c$, we thus have: $\rho(\zeta) = \frac{\rho_c}{L_c} \zeta$. We can also write $\frac{\rho_c}{L_c} = \tan\beta$, where $\beta$ is the cone's vertex angle. 
\par
Now, consider an infinitesimally wide circular band on the cone, of radius $\rho$, angular extent $\phi_c$, and width $\frac{d\zeta}{\cos \beta}$. The band has bending energy $dU = \frac{B}{2} \times \kappa_N^2 \times (\phi_c \rho) \times \frac{d\zeta}{\cos \beta}$, where $B$ is the bending modulus and $\kappa_N$ is the local normal (i.e.~out-of-plane) curvature. Given the geometry of the cone, we have $\kappa_N = \kappa \cos\beta$, where $\kappa=1/\rho$ is the arc curvature of the circular band. Thus $dU = \frac{B}{2} \times (\cos\beta/\rho)^2 \times (\phi_c \rho) \times \frac{d\zeta}{\cos \beta}$.  
Simplifying and integrating over an entire conical sector gives:
\begin{align}
	U_{\rm bend}^{\rm cone} &= \frac{B}{2}\phi_c \cos \beta \int_{0}^{L_c} d\zeta \; \frac{1}{\rho(\zeta)} \\
 &= \frac{B}{2} \phi_c \frac{\cos\beta}{ \tan\beta} \int_{0}^{L_c} \frac{d\zeta}{\zeta}
\end{align}
The singularity as $\zeta \to 0$ means that the conical shape must be modified there. Thus we consider only the region beyond some $L_{\rm core}$.  
\begin{equation}\label{eq:SI_UbendCone}
	U_{\rm bend}^{\rm cone} = \frac{B}{2} \phi_c \frac{\cos\beta^2}{\sin \beta} \log (\frac{L_c}{L_{\rm core}}),
\end{equation}
where $ \beta = \tan^{-1} \frac{\rho_c}{L_c}$. 
For the majority of numerical samples discussed in this paper, where $m \gg \rm{max} \{1,1/w\} $, it is sufficient to use small-angle approximations for $\beta$.
Thus, setting $\cos \beta \approx 1$ and $\sin \beta \approx \beta$, Eq.~(\ref{eq:SI_UbendCone}) gets reduced to :
\begin{equation}
	U_{\rm bend}^{\rm cone} \approx \frac{B}{2} \phi_c \frac{L_c}{\rho_c} \log (\frac{L_c}{L_{\rm core}}),
\end{equation}
Here, $\phi_c$, $\rho_c$ and $L_c$ are all functions of the contraction $\Delta$. But $\phi_c$ and $\rho_c$ are related through the constraint: $\phi_c \rho_c = L = \frac{\pi (1+w)}{m}$ (see Eq.~\ref{eq:SI_coneConstruct_inextensibility}), and $L_c \approx w$ (the width of the annulus), so $U_{\rm bend}^{\rm cone}$ can be reduced to a function of the single dynamic variable $\rho_c (\Delta)$. For the entire annulus, we need to multiply this by the number of cones $2m$. Thus, we get:
\begin{equation}\label{eq:SI_UbendCone_final}
	U_{\rm bend}^{\rm cone} \approx B \frac{\pi (1+w) w}{\rho_c^2(\Delta)} \log (\frac{w}{L_{\rm core}}),
\end{equation}
More generally, for a cone extending between axial limits $L_{\rm min}$ and $L_{\rm max}$, we have: 
\begin{equation}\label{eq:SI_UbendCone_final_generalLimits}
 U_{\rm bend}^{\rm cone} (\Delta) \approx B \frac{ \pi (1+w) w}{\rho_c^2(\Delta)} \log (\frac{L_{\rm max}}{L_{\rm min}}),
\end{equation}
This is the expression given in the main text. 
\par
While Eqs.~(\ref{eq:SI_UbendCone_final}) and (\ref{eq:SI_UbendCone_final_generalLimits}) seem independent of $m$, it is not so. Eq.~(\ref{eq:rhoc}) in the main text shows that $\rho_c \sim (1+w)/m \implies U_{\rm bend}^{\rm theory} \sim m^2$, as expected of a bending energy. We note that this $\rho_c \sim 1/m$ scaling could have been predicted in another way. In the limit of maximum possible contraction ($\Delta \to 1$), the point C approaches the xy-plane, and so the diameters of the $2m$ circles must approximately equal the reduced projected outer perimeter: $2\pi (1+w-\Delta )\to 2\pi w$. This gives us: $ 4m \rho_c \to 2 \pi w \implies \rho_c \to \frac{\pi w}{2m} $ (the dotted line in the inset of SI Fig.~\ref{fig:SI_fig2}c). Indeed, even if C is off the x-y plane (e.g.~for smaller $\Delta)$, the diameters of the $2m$ circles must still equal the reduced perimeter $2\pi(1+w-\Delta)$ up to some factor. Thus, we have the scaling relation $\rho_c \sim 1/m$ as expected.
\par
Finally, we note that the data presented in the paper represents the full expression \ref{eq:SI_UbendCone}, without any approximation.
\subsection{Limits of the cone-triangle construction} \label{sec:SI_construction_limits}
\begin{figure}[htb]
\centering
\includegraphics[width=0.48\textwidth]{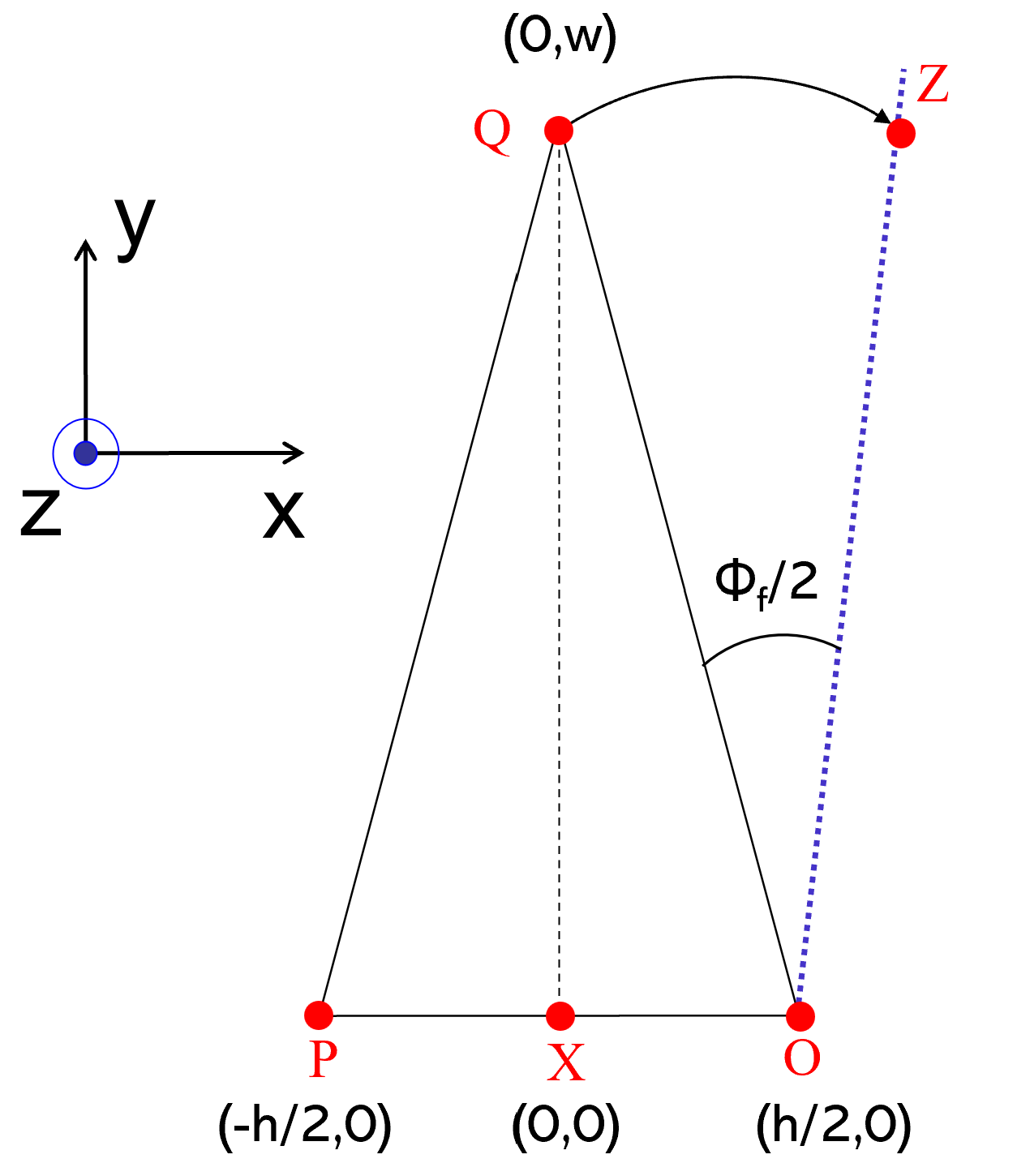}
\caption{Two different flat state geometries for $w=0.2$. (Left) For $m=4$, the flat angle $\phi_f < \pi$, which allows the conical construction described in Sec.~\ref{sec:SI_construction_limits}. (Right) For $m=2$, the flat angle $\phi_f > \pi$, which means that our conical construction is not valid here.}
\label{fig:SI_fig4}
\end{figure}

To the best of our knowledge, the above conical construction works as long as the initial (flat) angle $\phi_f < \pi$, i.e.~as long as the edges of two adjacent triangles define a triangle. The value of $\phi_f$ depends on the width $w$ and the wavenumber $m$, and increases as $m$ decreases. This defines a minimum wavenumber $m_{\rm min}^{\rm cone} (w)$ below which a conical solution is invalid. For a sufficiently wide annulus, we find that $m_{\rm min}^{\rm cone} =2$, which is the minimum possible value for \textit{any} wrinkled solution. However, for very narrow annuli, this value goes up. Thus, for $w=0.2$, we find $m_{\rm min}^{\rm cone} = 4$. Fig.~\ref{fig:SI_fig4} shows two such contrasting geometries. $m_{\rm min}^{\rm cone} (w)$ defines a geometric limit beyond which we expect our cone-triangle model to fail. However, the main text shows that we already see significant deviations from our model for Abaqus solutions with wavenumber $m$ significantly higher than $m_{\rm min}^{\rm cone} (w)$.

\subsection{ Comparison with separable, sinusoidal wrinkling} 
\label{SI:sinusoidal}
The cone-triangle shape described in the main text sharply reduces the elastic strain and energy relative to conventional wrinkling with height profile $h^c(r, \theta)$ {\em e.g.} of the form \cite{Davidovitch2019}
\begin{equation}
h^c(r, \theta) = f(r) \cos(m \theta)
\end{equation}
as we now illustrate.

In the far-from-threshold regime, this $h^c$ profile must relax the azimuthal strain $\epsilon_{\theta\theta}$ arising from the inward displacement $u_r = -\Delta$. In general this azimuthal strain has the form \cite{Davidovitch2011}: 
\begin{equation}\label{eq:SI_epsilontt}
\epsilon_{\theta \theta} = {u_r \over r} + {1 \over r} \partial_\theta u_\theta + {1 \over 2r^2} (\partial_\theta h^c)^2 
\end{equation}
For purposes of comparison we may evaluate this $\epsilon$ at a zero of $h^c(r, \theta)$ {\em e.g.} $m\theta = \pi/2$. There by symmetry the azimuthal displacement $u_\theta = 0$ and $\partial_\theta u_\theta = 0$. 
Thus the second term in Eq. \ref{eq:SI_epsilontt} vanishes. In the last term $\partial_\theta h^c = \pm m f(r)$. We use this expression to estimate $\epsilon_{\theta\theta}$. Choosing $h^c(r, \theta)$ to make $\epsilon_{\theta\theta}$ vanish implies
\begin{equation}
\Delta \simeq {1\over 2r} m^2 f(r){}^2
\end{equation}
So that $f(r) \simeq \frac{1}{m} \sqrt{2 r \Delta}$

This non-constant $f(r)$ entails a radial strain $\epsilon_{rr}$ given by \cite{Davidovitch2011}
\begin{equation}
\epsilon_{rr} = \partial_r u_r + \frac12 (\partial_r h)^2
\end{equation}
Here $u_r$ is constant as noted above and $\partial_r h\simeq \partial_r f(r) \simeq \frac{1}{m} \frac12 \sqrt{2\Delta/r}$, giving an estimated conventional wrinkle strain $\epsilon_{rr}^c$ of 
\begin{equation}
\epsilon_{rr}^c \simeq \frac{1}{4} m^{-2} (\Delta/r)
\end{equation}

This radial strain of conventional wrinkling is to be compared to the bending strain $\epsilon^B$ in the cone-triangle model. This strain arises from the bending stress $\sigma_{\theta\theta}^B$ \cite{Davidovitch2019}
\begin{equation}\label{eq:SI_sigmaB}
|\sigma_{\theta\theta}^B | \simeq 2 B (m/2\pi r)^2 , 
\end{equation}
where bending modulus $B$ is related to the thickness $t$, the Poisson ratio $\nu$ and the bulk Young's modulus $E$ by \cite{Landau:1986sf} Sec.12
\begin{equation}
B = Et {t^2 \over 12(1-\nu^2)}.
\end{equation}
Using $\epsilon = (Et) \sigma$, then Eq. \ref{eq:SI_sigmaB} yields
\begin{equation}
\epsilon_{\theta\theta}^B \simeq 2 {t^2 \over 12(1-\nu^2)} (m/2\pi r)^2 
\end{equation}
For the annuli simulated above the conventional wrinkles have much greater elastic strain than the bending strain we report: their ratio is given by 
\begin{equation}
\frac{\epsilon^c}{\epsilon^B} \simeq m^{-4} (r\Delta /t^2) ~ \left(6(1-\nu^2)\pi^2 \right)
\end{equation}
Evidently for fixed $m$ the ratio diverges as $t\to 0$. For the specific annulus used for Fig 5 with $r = 1$, $\Delta = 0.2$, $\epsilon^c/\epsilon^B \simeq 50$.
By these estimates the sinusoidal wrinkling carries substantially larger strain and thence elastic energy relative to the isometric wrinkling we report. 

\begin{figure*}[htb]
\centering

\textbf{a)}\includegraphics[width=0.4\textwidth]{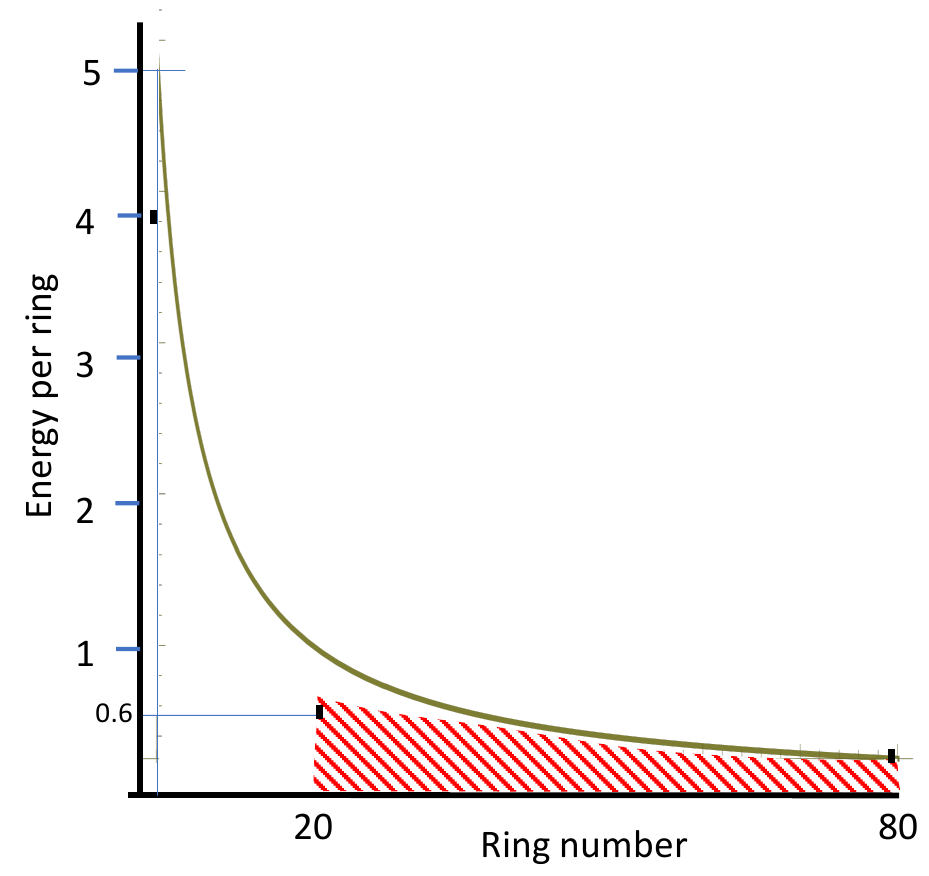}\quad\textbf{b)}\includegraphics[width=0.5\textwidth]{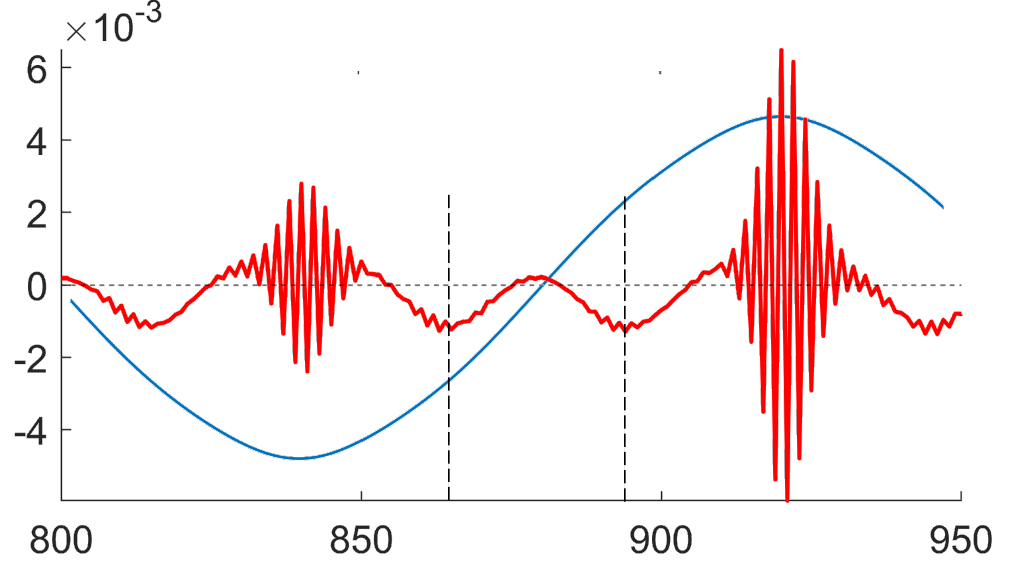}
\caption{a) Energy distributions for the annulus of Fig.~\ref{fig:5}. Horizontal axis is radial distance from inner rim measured in finite element widths. Vertical axis is ring energy described in the text. Hashed region shows the region treated in Fig.~\ref{fig:5}. Its area represents the simulated energy with largest $\Delta$ on the lower curve of Fig.~\ref{fig:5}. Black marks show simulated ring energies at three selected radii. Solid curve shows the ring energy profile calculated from cone-triangle model used for the middle curve in Fig.~\ref{fig:5}. b) Simulated azimuthal strain $\epsilon_{\theta\theta}$ (in red) as a function of azimuthal finite-element coordinate $\theta$, for a material circle located at the middle of an annulus with ($w=0.33$, $t=1.33 \times 10^{-3}$, $m=10$, $\Delta=0.27$). Light curve in blue shows height profile. Region spanning one trough and peak are shown. Variability of other troughs and peaks is similar. Vertical black lines indicate the location of two consecutive cone-triangle boundaries. 
}
\label{fig:SI_fig5}
\end{figure*}

\subsection{Departures from conical shape}
\label{sec:departures}

Here we gauge the impact of the stretched zigzag vertices on the total energy by examining how the elastic energy varies with distance $r-1$ from the inner rim. Fig.~\ref{fig:SI_fig5}a shows this energy for the right-most data points of Fig.~\ref{fig:5}. The red simulation point (lower curve) in Fig.~\ref{fig:5} is found by summing the energies in each azimuthal ring of finite elements over the range of $r$ indicated by the hashed region. The simulated ring energies for selected $r$ are shown as black rectangles. 

The corresponding cone-triangle energy, \textit{i.e.~}the right-most point on the (yellow) middle curve in Fig.~\ref{fig:5}, is found using the cone ring energy plotted as a solid curve, using Eq.~\ref{eq:Ubend_cone_kappa}. This equation implies a ring energy varying as $1/(r-1)$. Since the yellow point is found by matching the simulated curvature at the outer boundary, the cone model gives a ring energy that matches the simulated ring energy there, as Fig.~\ref{fig:SI_fig5}a shows. 

By comparing the simulated ring energies with the cone model, we can gain insight into the difference in energies seen in Fig.~\ref{fig:5}. The inner simulation point shows the expected large discrepancy with the model, which unrealistically diverges at the inner boundary. The middle simulation point is at the inner boundary of the region treated in Fig.~\ref{fig:5}. This ring energy is 40 percent smaller than the cone model prediction. This difference is consistent with the 15 percent differences between the energies of Fig.~\ref{fig:5}. Any cone that would cure this discrepancy would have to extrapolate to a vertex beyond the inner boundary.

The azimuthal strain profile shown in Fig.~\ref{fig:SI_fig5}b also shows departures from the model. The plot shows the azimuthal dependence of the azimuthal strain $\epsilon_{\theta \theta}$ for a radial position $r$ equidistant between the inner and outer boundaries.. The plotted strain is maximal at peaks and troughs of the wrinkles. Though the averaged strains,are consistent with bending strains, as discussed in Sec.~\ref{sec:results_inextensibility}, there are strong oscillations adjacent to these peaks and troughs with a period of two finite elements. This suggests that our simulations have limited reliability for predicting these weak strains near the peaks or troughs. Away from the peaks and troughs the strains vary smoothly. Of special interest is the point where the model cone and its adjacent triangle would meet. These points, marked by dashed lines show no sign of discontinuity.

These detailed features of Fig.~\ref{fig:SI_fig5} show shortcomings of the cone-triangle model. However they underscore the relevance of this model for understanding this form of buckling. 


\subsection{Finite-element method (FEM) simulation details}
For our simulations, we used the commercial finite-element software Abaqus 2018 (Simulia, Dassault-Systèmes, Providence, RI). This section describes the different steps for generating a typical simulation of our inner Lamé system, in the order typical of a finite-element software.
\par
The assembly consisted of only a single annulus, with inner radius fixed and taken to be unity, and with varying width $w$ and thickness $t$ in order to test our system over a wide range of system parameters. For width, we used values $w=0.20, 0.33, 0.67,1.0, 1.67$ (a factor of almost 10, ranging from very narrow to moderately wide), and for thickness, we used values $t=2.67\times10^{-3}, 1.33\times10^{-3}, 6.67\times10^{-4}, 2.67\times10^{-4}, 1.33\times10^{-4}$ (a factor of 20, ranging from moderately thick to very thin). While these thickness values vary over a decade, the values still fall well within the thin sheet limit. The annular part was made of 2d shell quad (S4R) elements \cite{Abaqus2018}. This choice was made mainly to optimise speed, since we used a fine enough mesh to ensure that doubling the linear mesh size change the energy by a negligible amount ($\lessapprox 1-2\%$). For comparison, the coarsest mesh we used was for the $w=1.67$ annulus, with 60 elements across the radius and 1400 elements across a circle, giving a maximum linear size for an element $\approx 0.01$. For consistency checks however, we also ran some simulations with annuli made of 3d volume cubic (C3D8R) elements \cite{Abaqus2018}, which gave the same morphology (with the same wavenumber), but which much longer running times.
\par
When discussing the material properties, for concreteness, we will use SI units (and thus take the inner radius to be $1$ m). For the material properties, we mostly used a standard neo-Hookean hyperelastic model \cite{Abaqus2018} with coefficients $\rm C_{10} = 1.5375\times 10^5 \; \rm Pa$, $\rm D_1 = 3.2520\times 10^{-7} \; \rm Pa^{-1}$. These coefficients are related to the more well-known linear elasticity moduli by the relations: $C_{10} = \rm G/2$ (where $\rm G$ is the shear modulus), and $\rm D_1 = 2/K$ (where $\rm K $ is the material bulk modulus). The corresponding Poisson ratio is given by $\nu = \frac{3/(C_{10} D_1)-2}{6/(C_{10} D_1)+2} =0.475$. The Young's modulus can be obtained from either of the relations: $E = 2G (1 + \nu),$ or $E = 3K (1 - 2 \nu)$; we obtain $E = 907,377$ Pa $\approx 0.9 $ MPa (corresponding to a rubber-like material). As a test, we also performed several simulations with a linear material model with these values of $E$ and $\nu$. 
\par
The elastic modulus can be used, along with the material density $\rho$ and the average linear mesh size $l_e$, to determine the average integration time scale (i.e.~the `stable time increment') in the simulation, as follows. The elastic bulk modulus $K$ and the density $\rho$ determine the speed of sound in the material, $c_s = \sqrt{K/\rho}$ \footnote{Alternately, one can use the Young's modulus $E$ instead of the bulk modulus $K$ in the definition of $c_s$. But this does not qualitatively alter our argument above.}. The stable time increment in the simulation is on the scale of the time required for elastic information to traverse an average mesh element: $\Delta t \sim l_e / c_s$. For our values of $K \sim 10^5$ Pa, $\rho = 10^3 \; \rm Kg/m^3$ and $l_e \sim 10^{-2}$ m, we get $\Delta t \sim 10^{-3}$ s. This in turn determines the dimensionless number of iterations $n_{\rm iter}$ performed by the solver in a simulation running over time period $T$: $n_{\rm iter} = T/\Delta t$. Below, we discuss the typical values of $T$ used in our simulations, and how increasing $T$ allows us to reach a quasi-static limit in dynamic integration methods (i.e.~where kinetic energy is present but negligible compare to elastic energy). 

\par
For the radial displacement loading at the inner boundary, we applied velocity and displacement boundary conditions (BCs) interchangeably. Typically, we applied velocity BCs with a linear amplitude profile, and displacement BCs with a smooth-step profile \cite{Abaqus2018}, in order to assure a smooth (i.e.~zero velocity) pull at the beginning. These were applied so that the maximum displacement amplitude $\Delta_{\rm max} = 0.267$ is attained within a time period $T$ (defined in the same units as the $\Delta t$ given above). The value of $T$ was chosen to be large enough to ensure small kinetic energy and give a $T$-independent configuration, as defined below. Typically, we used $T=20$ for the thicker sheets, and $T=80$ for the thinnest sheets.
\par
For the simulation protocol, we employed both `dynamic explicit' and `dynamic implicit' \cite{Abaqus2018} integration schemes in the quasi-static limit (as compared to a fully `static' energy minimisation scheme). The word `dynamic' refers to the presence of inertia, while `explicit' and `implicit' refer to the solution scheme. `Explicit' means explicit time-integration of Newton's second law, while `implicit' refers to implicit integration (viz.~through iterative root-finding) of Newton's law, using a modified Newton-Raphson method. 
The mixture of these two methods was done partly as a consistency check, partly for convenience, and partly by necessity. While the implicit method in the quasi-static limit was faster for most jobs, the explicit solver was indispensable for the thinnest samples, where the static solver ran into convergence problems. Ensuring the quasi-static limit is also easier for `dynamic implicit' than for `dynamic explicit'. In dynamic implicit, the quasi-static option is in-built, but for dynamic explicit, it has to be ensured manually by applying the loading slowly enough so that further slowing has no effect on the final shape and energy. 
\par
For this, we need to look at the available energy modes. The energy balance equation in Abaqus \cite{Abaqus2018} is given by (ignoring possible terms coming from viscosity, friction, heat, contact and constraint penalties, etc.), is:
\begin{equation}\label{eq:SI_energyBalance}
 E_I + E_{KE} - E_W = 0
\end{equation}
 where $E_I$ is the internal energy, $E_{KE}$ is the kinetic energy, and $E_W$ is the work done by externally applied loads. For us, the internal energy is just the elastic energy (by design, there are no other energy modes). Thus, for quasi-static loading, one generally requires the kinetic energy ($E_{KE}$) to be $<10\%$ of the elastic energy ($E_I$). In practice, we kept the ratio to $\lessapprox 5\%$. Since the elastic energy is thickness-dependent (always increasing with increasing thickness), thinner sheets required slower applications of the loading.
E.g., for the thinnest samples ($t=1.33\times10^{-4}$), this involved applying the contraction $\Delta$ over a time period $T=80-120$, viz.~using $n_{\rm iter} \sim 10^5$ solver iterations. As a result, the slowest simulations, for the thinnest and widest sheets, lasted $\approx 150$ core-hours; the average simulation however, lasted between $\approx 30$ core-hours. For reference, the validation case (described blow) of folding a flat sheet into a cylinder, albeit with a much coarser mesh, was accomplished using $\approx 0.2$ core-hours. 
\par
For data extraction, we used field output for the displacement variables, and history output for the energy \cite{Abaqus2018}. For the elastic energy, we used the ``ALLIE" (internal energy) variable \cite{Abaqus2018}, equivalent to the $E_I$ variable in Eq.~\ref{eq:SI_energyBalance}. Since the simulations are done using a dynamic time-integration scheme with inertia, there is an inherent noise in the energy values arising from imprecision in finding the exact energy minimum (viz.~due to inertial oscillations). We cannot estimate this noise precisely, but a rule-of-thumb estimate is $\lessapprox 5\%$, i.e.~of the same order of magnitude as the ratio $E_{KE}/E_I$. However, in reality, it might well be less. Significantly, this noise cannot account for the discrepancy between energy measurements and model predictions in Fig.~of the main text. Finally, post-processing was done using Abaqus2Matlab \cite{Papazafeiropoulos2017}.

\subsection{Testing the numerics for known cases}
We used the above procedures (albeit with a static energy minimisation scheme) to calculate an analytically solvable case, to verify that the shape and energy agreed with the known results. The example was a rectangular sheet of width $w=1$, length $L=2\pi$, and thickness $t=1\times10^{-3}$, in which we prescribed boundary conditions on position and orientation of the short edges to make them curve up and in, to form a circular cylinder of unit radius. 
\par
We verified the circularity of the cross-section by projecting onto the plane and measuring the distance from the centre. We found that no point differed in its axial distance by more than $.001\%$. The measured elastic energy differed only slightly from the analytic result, $U^{\rm cylinder}_{\rm bend} = B \pi \approx 3.07 \times 10^{-4}$ Joule, where $B= E t^3/12 (1-\nu^2)$ is the bending modulus, obtained using the thickness $t$, Young modulus $E$ and Poisson ratio $\nu$ quoted above. The simulation gave an energy $ \approx .01\%$ larger than this. A discrepancy of this sign is expected because the analytic form neglects the small strain energy owing to the nonzero thickness of the sheet simulated.

---------------------------------------------------

\end{document}